\documentstyle[epsf]{article}

\newcommand{\be}{\begin{equation}}
\newcommand{\ee}{\end{equation}}
\newcommand{\bea}{\begin{eqnarray}}
\newcommand{\eea}{\end{eqnarray}}
\newcommand{\beas}{\begin{eqnarray*}}
\newcommand{\eeas}{\end{eqnarray*}}
\newcommand{\bdm}{\begin{displaymath}}
\newcommand{\edm}{\end{displaymath}}
\newcommand{\ba}{\begin{array}}
\newcommand{\ea}{\end{array}}
\newcommand{\bi}{\begin{itemize}}
\newcommand{\ei}{\end{itemize}}
\newcommand{\ben}{\begin{enumerate}}
\newcommand{\een}{\end{enumerate}}
\newcommand{\bc}{\begin{center}}
\newcommand{\ec}{\end{center}}
\newcommand{\bfl}{\begin{flushleft}}
\newcommand{\efl}{\end{flushleft}}
\newcommand{\bfr}{\begin{flushright}}
\newcommand{\efr}{\end{flushright}}
\newcommand{\bd}{\begin{description}}
\newcommand{\ed}{\end{description}}
\newcommand{\bq}{\begin{quote}}
\newcommand{\eq}{\end{quote}}
\newcommand{\bfg}{\begin{figure}}
\newcommand{\efg}{\end{figure}}
\newcommand{\bt}{\begin{table}}
\newcommand{\et}{\end{table}}
\newcommand{\btb}{\begin{tabular}}
\newcommand{\etb}{\end{tabular}}
\newcommand{\btg}{\begin{tabbing}}
\newcommand{\etg}{\end{tabbing}}

\renewcommand{\Im}{\mbox{Im} \,}

\newcommand{\kslash}
           {\mbox{$ k \hspace{-1.1ex} \mbox{/} \hspace{-0.07ex} $}}

\newcommand{\sslash}
           {\mbox{$ s \hspace{-1.1ex} \mbox{/} \hspace{-0.07ex} $}}

\begin{document}
\newtheorem{theorem}{Theorem}
\newtheorem{prop}{Proposition}
\newtheorem{lemma}{Lemma}
\newtheorem{defi}{Definition}
\renewcommand{\thefootnote}{\fnsymbol{footnote}}
\noindent April 1997\hfill
${\mbox{MZ-TH/96-36}\atop \mbox{hep-th/9612010}}$\\[1cm]
\begin{center}
{\Large \bf Weight Systems from Feynman Diagrams\\[1cm]}
{\bf Dirk Kreimer\footnote{supported by DFG}\footnote{email:
kreimer@dipmza.physik.uni-mainz.de}\\}
{\it Dept.\ of Physics\\
Mainz University\\
55099 Mainz\\
Germany\\[1cm]}
\end{center}
\renewcommand{\thefootnote}{\arabic{footnote}}
\setcounter{footnote}{0}
\begin{abstract}
We find that the overall UV divergences of a renormalizable
field theory with trivalent vertices fulfil a four-term relation.
They thus come close to establish a weight system.
This provides a first explanation of the recent successful association
of renormalization theory with knot theory.
\end{abstract}
\section{Introduction}
Recently, ideas spread to connect knot theory with renormalization theory
\cite{DK1,DK2}.
This was guided by the observation that overall counterterms
are independent of external parameters like masses and momenta.
This well-known fact can be cast in a different form: that
the only way to change the value of an overall divergence
is to change the topology of the diagram,
once the spin and other representations
 of the involved particles are specified.
Hence one should not be too surprised if one
finds topological information in the numbers which specify
the overall divergences. And indeed, meanwhile evidence
in abundance reports on a beautiful and unexpected connection
between field theory \cite{BKP,BDK,BGK,DKT,IMU3},
knot theory \cite{DK3,BK15} and number theory \cite{EUL,BG,BBB,TERA}.

On the one hand these results successfully explore the proposed
connection
by empirical calculations, while on the other hand an explanation why it
works is missing. The connection between
field theory and number theory is given by solid calculus,
the connection between field theory and knot theory via
momentum routings  is fairly
intuitive \cite{DK1}, and
finally the connection between knot and number theory, which the
results in \cite{BKP,BGK,DK3,BK15,EUL} so urgently
suggest, is still to be found.

In this paper, we will study four-term relations.
Such relations appear in a totally different context, related to
knot theory via Vassiliev invariants and their study
in terms of chord diagrams \cite{Dror}.

Here we want to derive the presence of a four-term relation
in divergent contributions extracted from Feynman diagrams.
In a companion paper, we will present calculations
which provide examples for the ideas proposed here \cite{BK4}.

Such a four-term relation (4TR) is a relation between trivalent
graphs. It is thus perfectly well suited to help
us with our main problem: how to relate trivalent Feynman graphs
to knot diagrams.

The presence of a 4TR provides an algebraic structure
which directly relates overall counterterms of Feynman diagrams
to knot theory. This, we hope, will provide the first and major step
in a formal derivation of the relation between renormalization
theory and knot theory. The 4TR relation is intimately connected
to chord diagrams. Such chord diagrams were envisaged
to play a role for our purposes in \cite{DK1} and were also
investigated in \cite{JKTT}. While we disagree
with the prescriptions used in \cite{JKTT}, since they quickly
lead to violations of established connections between knots
and transcendentals from counterterms \cite{DK1}, the intent
of that work formed some of the motivation for the present analysis.

The paper is organized as follows. In the next section we will
motivate the four-term relation and the role it plays
in the study of knot invariants of finite type. Our presentation
follows \cite{Dror} closely.
Having convinced ourselves that the
presence of a four-term relation is the major ingredient
to establish algebraic structures which relate to knot theory,
we then derive in the main section
the presence of a 4TR in the counterterms
of overall divergent diagrams in a renormalizable field theory.
In the final section we consider a curious example, extending the
consideration to the presence of non-renormalizable couplings.
We derive a relation, which we will report to be fulfilled in
a different paper \cite{BK4}.\footnote{Actually, this relation
was postdicted after David Broadhurst proved the
4TR to fail in its naive form in graphs with
non-renormalizable vertices, and I was forced to explain this
failure.}

In this first approach to the subject, we will restrict ourselves
to graphs free of subdivergences. We do expect that the general
case will be a modification of the results obtained here, but will
reserve a more general  study for the future.

\section{The four-term relation}
Let us first define a chord diagram. Strictly following \cite{Dror},
we define:
\begin{defi}
A chord diagram is an oriented circle with finitely many chords on it,
regarded up to orientation preserving diffeomorphisms of the circle.
\end{defi}
Here, a chord is a pair of distinct points on the circle,
connected by a propagator. In \cite{Dror}, one then continues
to consider the class of all chords, filtered in a natural
manner by the number of chords (which is one less than
the number of loops in the diagram).
Adopting the notation of
\cite{Dror}, we call the set of all diagrams having $m$ chords
${\cal G}_m{\cal D}^c$.

One of the fundamental results in \cite{Dror} is the fact that the vector
space of Vassiliev invariants is equivalent to the graded space
of all weight systems. A weight system is more or less determined
by the existence of a 4TR in chord diagrams:
\begin{figure}
\epsfysize=5cm \epsfbox{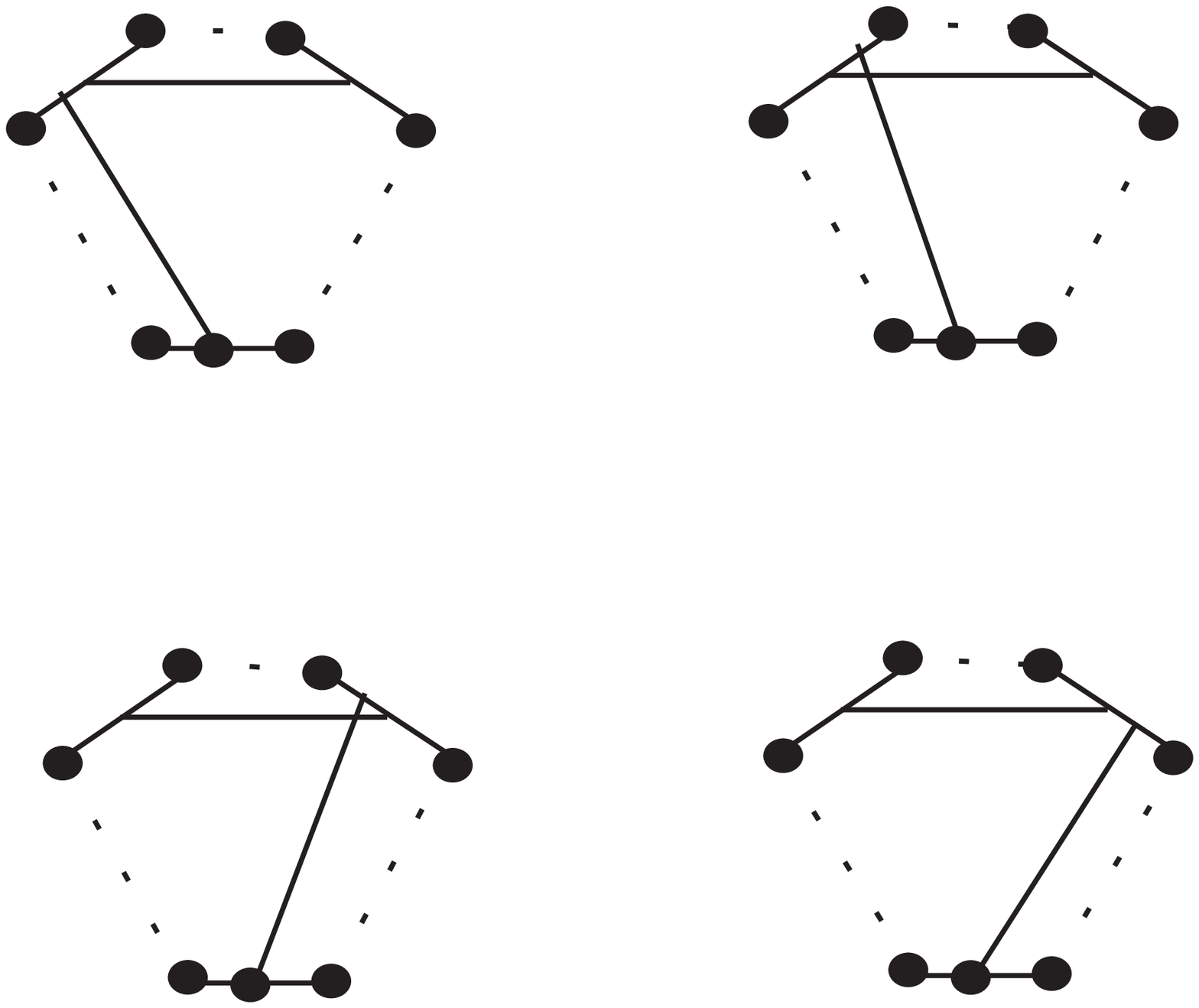}
\caption{\label{fig4tr}The four-term relation.
The alternating sum of the four diagrams ${\cal G}_{1,\ldots 4}$
vanishes.
Note that lines solely specify the topology of the diagram,
and not the nature of the particles involved. Dotted lines may be connected
in arbitrary ways by further propagators in the
same manner for all four diagrams, with the restriction that
all four graphs are free of subdivergences.}
\end{figure}
\begin{defi}
A real-valued weight system of degree $m$ is a function
$W$: ${\cal G}_m{\cal D}^c\to {\bf R}$ such that
\begin{description}
\item[i)] If $D$ $\in$ ${\cal G}_m{\cal D}^c$ has an isolated chord,
then $W(D)=0$.
\item[ii)] Whenever four diagrams $D_1,\ldots,D_4$
differ only as shown in Fig.(\ref{fig4tr}),
their weights satisfy
\begin{equation}
W(D_1)-W(D_2)+W(D_3)-W(D_4)=0.\;\;\mbox{(4TR)}
\end{equation}
\end{description}
\end{defi}
An isolated chord can be considered as a one-loop subdivergence
at the circle.

\begin{figure}[hb]
\epsfysize=3cm
\epsfbox{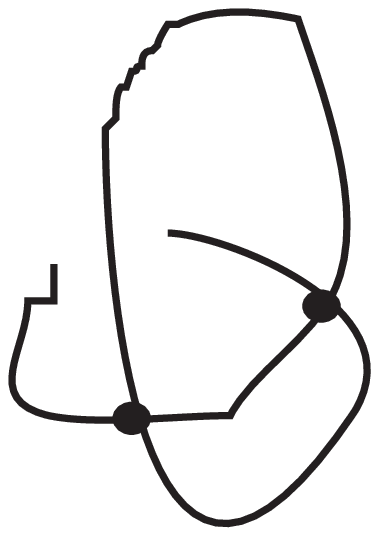}
\caption{\label{figknot}This is a knot with two selfintersections.
Any knot invariant $V$ can be lifted to a knot
invariant on knots with selfintersections.}
\end{figure}

For our purposes, property $i)$ is thus trivially fulfilled.\footnote{This
is consistent with the results in \cite{DK1}, which relate iterated
one-loop subdivergences to an increasing writhe number,
and thus establish a framing dependence. The absence
of subdivergences guarantees framing independence, which is
the sole purpose of property $i)$ in \cite{Dror}.}
Property $ii)$ is called the four term relation (4TR).
Let us denote by ${\cal W}$ the space of all weight systems.
Then the before-mentioned theorem says that ${\cal W}$
is equivalent to the filtered space of Vassiliev invariants over
${\bf R}$. To understand what this means let us consider knots
which have $m$ selfintersections, cf.~Fig.(\ref{figknot}).
Let us call the space of all knots with $m$ selfintersections ${\cal K}_m$
for the moment.

Assume some knot invariant $V$ is given. We then make it into
an invariant of knots with selfintersections by setting
\begin{equation}
V(\epsfysize=3mm\epsfbox{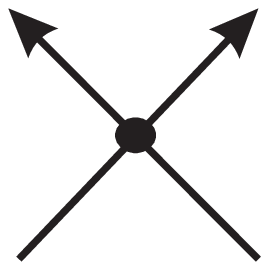})=
V(\epsfysize=3mm\epsfbox{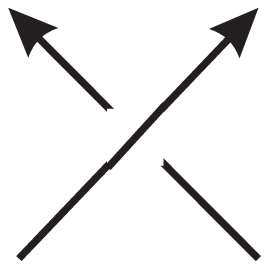})
-V(\epsfysize=3mm\epsfbox{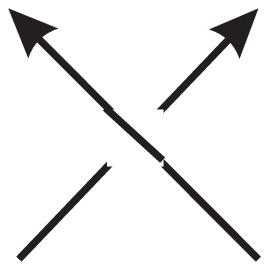}).
\end{equation}
Resolving all selfintersections with help of this equation iteratively
defines the knot invariant on ${\cal K}_m,\;\forall m$.
A knot invariant is called an invariant of type $m$ if it vanishes
for knots with more than $m$ selfintersections.
\begin{equation}
K\in\;{\cal K}_{i},\;i>m\;\Rightarrow\;V(K)=0.
\end{equation}
A knot invariant which is of finite type for some $m\in{\bf N}$
is called a Vassiliev invariant. The important point about Vassiliev
invariants is that all the coefficients of the relevant knot polynomials,
like the Conway, Jones, HOMFLY or Kauffman polynomials, are
invariants of finite type.

These results show the importance of a 4TR.
Once a mapping from trivalent graphs $G_i$ to ${\bf R}$ is
given which fulfils the 4TR on chord diagrams,
we can make it into a weight system by defining a new mapping
which vanishes on diagrams with isolated chords, and get from
the weight system a Vassiliev knot invariant.

The problem we address in this paper can be stated as follows:
counterterms of Feynman diagrams assign to each Feynman graph,
considered as a trivalent graph, a (Laurent-) series in a regularization
parameter with
coefficients in ${\bf R}$.\footnote{We are only interested
in the proper divergent part  of this series.}
For graphs free of subdivergences this is in fact a single number,
specifying the overall divergence of the graph.
For such graphs condition $i)$ in the
definition of a weight system is trivially fulfilled.
What remains to be checked is the 4TR.

%To get a feeling for the equivalence between Vassiliev invariants and
%weight systems, let us show that each Vassiliev invariant defines
%a weight system. The reverse case is much harder, and we refer the
%interested reader to \cite{Dror}.
%
%Assume a knot invariant of finite type $m$ is given.
%Consider an embedding $K_D$  of $D\in\;{\cal G}_M{\cal D}^c$, which is
%an immersion $\mbox{S}^1\to{\bf R}^3$ of the circle into
%${\bf R}^3$ whose only singularities (double points)
%are traversal selfintersections
%and which violates bijectivity only for endpoints of the
%same chord. Then two different
%embeddings can be transformed into each other by a sequence
%of flips; exchanges of over- and undercrossings.
%The value of $V(K_D)$ remains unchanged by such a flip,
%as the difference is given by a diagram with $m+1$ double points,
%on which $V$ is assumed to vanish..

Before we start with our investigation of this problem, we
mention some more facts valid for diagrams which fulfil
a 4TR. We stress some interesting results from \cite{Dror}
and point out which aspects might
be interesting for the calculation of Feynman diagrams.
A more detailed discussion will be contained in \cite{book}.
\begin{itemize}
\item
It can be shown that the space of all chord diagrams
fulfilling a 4TR is equivalent to the space of all
diagrams fulfilling a STU-relation, which is a
three-term relation. We shall attempt to investigate the
STU relation in the future.
If it can be established as well, arbitrary Feynman diagrams
can be reduced to Feynman diagrams which can be written as chord diagrams.
\item
The resulting algebra of diagrams
provides a naturally defined product and coproduct, which makes
it into a commutative and cocommutative Hopf algebra. Such algebraic
structures, once identified in overall divergent counterterms,
might turn out to be useful in future calculations.
\item It is reasonably conjectured that all weight systems
come from Lie algebras \cite{Dror}.
If this is true their algebraic relations
should be reflected in overall divergent counterterms. If this
in turn is wrong, our overall counterterms might provide the first example
of a weight system which does not come from Lie algebra.\footnote{At
this point, it might be interesting to know how big these spaces
of diagrams really are, for increasing loop number.
Interestingly, row 3, table 6.1 in \cite{Dror} gives numbers
which seem intimately related but not identical
to the number of distinct
irreducible MZV´s \cite{DZ}
found in overall counterterms
of graphs with appropriate  loop number \cite{BK15,book}.
To compare entries of table 1 in \cite{BK15} with \cite{Dror},
add in row $M_n$ the entry for odd $n$ to the (even) entry for
$n-1$,
starting from $n=5$. The first disagreements occur at $n=15$ and $n=19$.}
\end{itemize}

\section{Overall divergences and 4TR}
The idea which we want to pursue combines field theory and
topology. On the one hand, we learned that symbolic graphs which are related
by a 4TR  establish weight systems. On the other hand we know
from previous work that the overall counterterms of Feynman diagrams
are connected to knot theory. So far, this was obtained as an empirical
fact, by establishing a knot-to-number dictionary.

Now, we want to come closer to an explanation of this phenomenon
by establishing a 4TR between the overall divergences of Feynman
graphs. Our input are Feynman graphs which will be shown to be
related
by a 4TR. For a start, we exclude cases which contain subdivergences.
A short glance at the Dyson-Schwinger equations ensures that
this restricts ourselves to vertex corrections.
Further, as we work with a renormalizable (but not
super-renormalizable) theory, we will
conclude that the graphs to be considered have logarithmic
degree of divergence.

In such a case, we can calculate the overall divergence
by nullifying all external masses and momenta in all
propagators. We then assume that we can cut the diagram
at appropriate places, and find the overall divergence of a
$n$-loop graph as the finite part of a $(n-1)$-loop two-point
function.

In so doing, we assume that the cutting is legitimate.
This restricts ourselves to places where we can cut without
generating IR divergences. As we see below, the derivation
of the 4TR determines where we must be able to cut the graphs.
This restricts the class of available graphs. In future work we will
discuss modifications of the 4TR which can incorporate such cases as well.

Also, in a parallel paper, we will present an explicit 4-loop calculation
which establishes the 4TR as proposed here, and demonstrates its failure
in cases when our derivation is not applicable \cite{BK4}.

We now want to prove that Feynman diagrams fulfil a
four-term relation (4TR).
\begin{theorem}[4TR]
Four Feynman graphs ${\cal G}_i$, $i=1,\ldots,4$,
of logarithmic degree of divergence, which fulfil
\begin{itemize}
\item[i)] that they are free of subdivergences (skeleton diagrams),
\item[ii)] that they are related as in Fig.(\ref{fig4tr}),
\item[iii)]
that the propagators connecting $x_m$ to $x_i$ with vertices $V_i$
at $x_i$,
 $i\in \;\{l,r\}$,
fulfil $\Delta_F(s)V_i=V_i\Delta_F(s)$
for any constant vector $s$,\footnote{A vector boson coupling
to a fermion thus would not fulfil this condition,
for example: $\sslash \gamma_\mu-\gamma_\mu \sslash\not=0$.}
\item[iv)] that the propagators connecting $x_a-x_l,x_b-x_l,
x_c-x_r,x_d-x_r$ have no formfactor of tensor rank two,\footnote{
Thus, scalar particles, fermions and gauge bosons
in the Feynman gauge pose no problem. If we
have zero momentum couplings at a fermion line we might be
in trouble, see below. Constraints $iii)$ and $iv)$
may be relaxed when one allows for modifications
of the 4-term relation.}
\item[v)]
that they provide a correction to a dimensionless
coupling constant of a renormalizable theory,\footnote{In a
renormalizable theory with only trivalent couplings,
provisos $i)$ and $v)$ are not independent. We include $v)$
as we will extend our considerations to more general cases
in the course of this paper.}
\end{itemize}
fulfil
\begin{equation}
<{\cal G}_1-{\cal G}_2+{\cal G}_3-{\cal G}_4>=0.\label{4tr}
\end{equation}
\end{theorem}
To prove this theorem, we use contour integrals.
We start with contour integrals which vanish
by the residue theorem.
But, as we will argue, they
pick up the coefficients
of the overall divergence of the four Feynman diagrams above,
as their sole divergent contribution.

Our guiding philosophy is the following. Assume that all internal
momentum integrations are cut-off by an appropriate $\lambda$.
Then, all our contour integrals to be defined below exist.
We then collect all terms which will diverge for $\lambda\to \infty$
after addition of all the contours.
We will find that the remaining divergent contributions establish
the 4TR as envisaged in the theorem.

Hence we will shortly define ten contour integrals
$I_{i}$, $i=1,\ldots,10$, with ten contours $\Gamma_i$,
which fulfil
\begin{equation}
I_i=0,\;\forall i\in 1,\ldots,10;\label{e1}
\end{equation}
because the contours are closed rectangles,
and the functions to be integrated are analytic inside
these rectangles.

Each contour $\Gamma_i$
consists of four paths $\gamma_{i,j}$, $j=1,\ldots,4$.
The contributions from these individual paths are denoted
by $I_{i,j}$. We then show that
\begin{equation}
0  =
<\sum_{i=1}^{10}\sum_{j=1}^4 I_{i,j}>=<I_{1,2}+I_{5,2}+I_{6,2}+I_{10,2}>,
\label{sum}
\end{equation}
where the first equality follows from Eq.(\ref{e1}),
while the second equality  has to be proved.
Finally, we will find that
\begin{eqnarray}
<I_{1,2}> & = & <{\cal G}_{1}>,\nonumber\\
<I_{10,2}> & = & -<{\cal G}_{2}>,\nonumber\\
<I_{6,2}> & = & <{\cal G}_{3}>,\label{match}\\
<I_{5,2}> & = & -<{\cal G}_{4}>\nonumber,
\end{eqnarray}
which proves the theorem.
Here, $<\ldots>$ denotes projection
onto the divergent part of the expression in the brackets.
This projection can be done in various ways. For our purposes,
it is sufficient to cut the last momentum integration.
The brackets then project onto the coefficients of the singularity
present in the final momentum integration.

To proceed, we have to introduce some more notation.
First, note that for space-time points (coordinates) we will
use variables from the end of the alphabet: $x,y,z$,
where we might use alpha-numerical subscripts to denote more of
them, like in $x_0,x_a,\ldots,x_m$.
They are Lorentz vectors in space-time, where we usually
omit to give the Lorentz indices, but, if needed,
denote them  by greek sub(super-)scripts.
We use a four-dimensional Euclidean metric, supposing that Wick rotation
is performed {\em ab initio} in Eqs.(\ref{Gs}-\ref{Ge}) below.
Unit vectors will be denoted by
a caret, for example $\hat{y}_\mu:=y_\mu/\sqrt{y^2}$,
where $y^2=y_\mu y^\mu$. The modulus of a vector
is denoted by $\bar{y}:=\sqrt{y^2}$.

For momenta we will use letters from the middle of the alphabet,
$k,l,m,\ldots,r$.
Occasionally, we will use the notation
$k\cdot y=k_\parallel\bar{y}$, where $k_\parallel:=k\cdot\hat{y}$.
Here, $k\cdot y$ denotes the scalar product $k_\mu y^\mu$.

\begin{figure}
\epsfysize=4cm \epsfbox{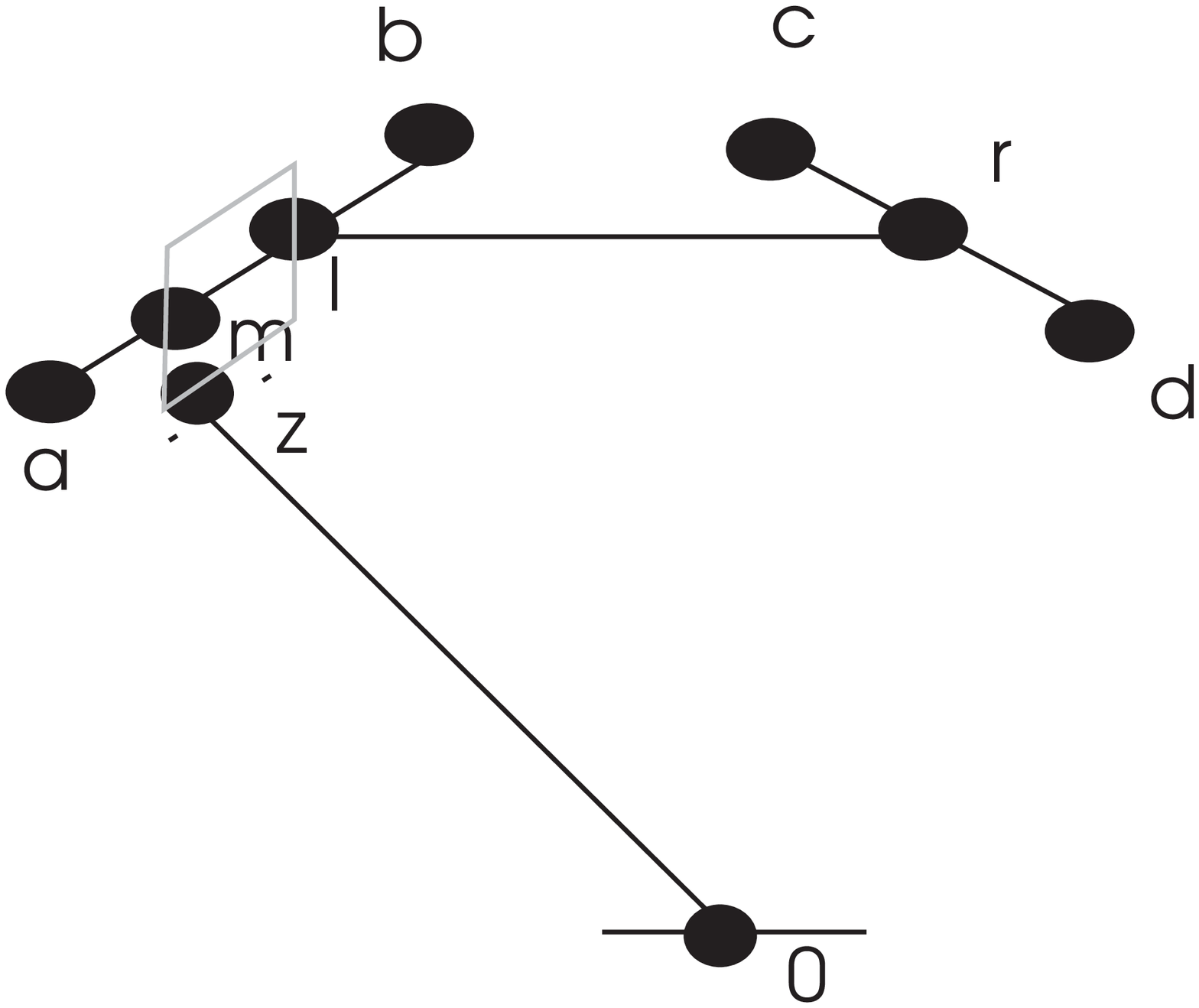}
\caption{\label{figz}The endpoint $z$ of the propagator
$\Delta_F(x_0-z)$ moves around a contour which lies in
the extension of the parallel space $x_m-x_l$ to a complex plane.
This figure refers to the case ${\cal G}_1$, where $x_m$ is
located between $x_a$ and $x_l$. Similar figures hold
for the other cases.}
\end{figure}

In Fig.(\ref{figz}) we now define the relevant propagators and vertices.
We specify 9 points: $x_a,x_b,x_c,x_d$, $x_0,x_l,x_r,x_m$ and $z$,
as indicated in the Fig.(3).
All differences between the four graphs of Fig.(\ref{fig4tr}) can be
specified by the propagators which connect the
points $(x_0,x_m,x_a,x_b,x_c,x_d,x_l,x_r)$.

The propagators of Fig.(\ref{figz}) correspond to one part of the analytic expression
for the diagrams ${\cal G}_i$. The rest is contained in a function
${\cal G}(x_a,x_b,x_c,x_d,x_0)$, which is universal for all
four diagrams.

These four diagrams ${\cal G}_i$ can now be written as\footnote{For
simplicity, we drop all factors coming from vertices.
We will comment on vertices later.}
\begin{eqnarray}
{\cal G}_1 & := &
\int dX {\cal G}(x_a,x_b,x_c,x_d,x_0)
\Delta_F(x_r-x_l)\Delta_F(x_0-x_m)\nonumber\\
 & & \Delta_F(x_m-x_a)\Delta_F(x_l-x_m)\Delta_F(x_b-x_l)
\Delta_F(x_r-x_c)\Delta_F(x_d-x_r),\label{Gs}\\
{\cal G}_2 & := &
\int dX {\cal G}(x_a,x_b,x_c,x_d,x_0)
\Delta_F(x_r-x_l)\Delta_F(x_0-x_m)\nonumber\\
 & & \Delta_F(x_l-x_a)\Delta_F(x_m-x_l)\Delta_F(x_b-x_m)
\Delta_F(x_r-x_c)\Delta_F(x_d-x_r),\\
{\cal G}_3 & := &
\int dX {\cal G}(x_a,x_b,x_c,x_d,x_0)
\Delta_F(x_r-x_l)\Delta_F(x_0-x_m)\nonumber\\
 & & \Delta_F(x_l-x_a)\Delta_F(x_b-x_l)\Delta_F(x_m-x_c)
\Delta_F(x_r-x_m)\Delta_F(x_d-x_r),\\
{\cal G}_4 & := &
\int dX {\cal G}(x_a,x_b,x_c,x_d,x_0)
\Delta_F(x_r-x_l)\Delta_F(x_0-x_m)\nonumber\\
 & & \Delta_F(x_l-x_a)\Delta_F(x_b-x_l)\Delta_F(x_r-x_c)
\Delta_F(x_m-x_r)\Delta_F(x_d-x_m).
\label{Ge}
\end{eqnarray}
In all four  ${\cal G}$ functions the propagator
$\Delta_F(x_0-x_m)$ appears. It will be the endpoint
$x_m$ of this propagator which will be the subject of a contour
integration.

${\cal G}(x_a,x_b,x_c,x_d,x_0)$
incorporates the five-point function which corresponds
to all other not specified propagators and vertices in the
Feynman diagrams. It is the same function in all four
cases. $\int dX$ summarizes the integration over all internal
vertices, $\int d^4x_a\ldots \int d^4x_m$. There might be other
internal vertices in ${\cal G}(x_a,x_b,x_c,x_d,x_0)$,
whose integration is always understood. We do not specify
where external particles couple, but in principle allow them to couple
anywhere. As we are interested only in the overall divergence,
we nullify masses and external particles in all propagators.
As mentioned before,
we assume that this can be done without
introducing IR problems.

We also use the Fourier transform $\tilde{\Delta}_F(k)$
\begin{equation}
\Delta_F(y)=:\int d^4k \tilde{\Delta}_F(k)
e^{ik\cdot y}.\label{fourier}
\end{equation}

\begin{figure}
\epsfysize=5cm \epsfbox{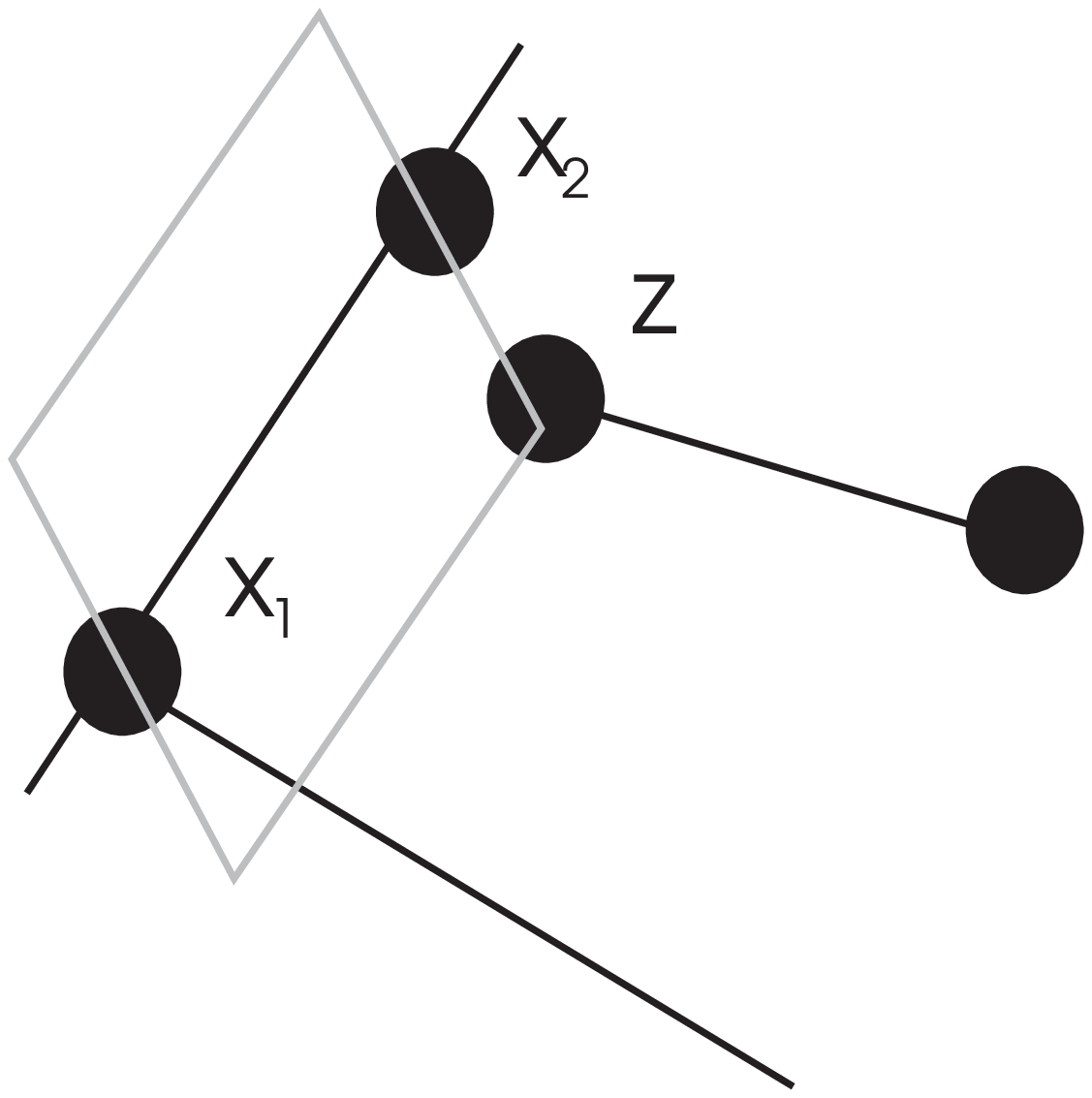}
\caption{\label{figf}The generic case for all our contour integrals.
Here, $y=x_1-x_2$.}
\end{figure}

Now we introduce the contour integrations which we have in mind.
We use
\begin{equation}
\oint_{\Gamma(y)}dz e^{iz k\cdot \hat{y}}=0,\;\;\forall k,y.\label{cont}
\end{equation}
The contour $\Gamma(y)$ is a contour
in a complex plane. The real axis of this complex plane
is identified as a multiple
of the vector $\hat{y}_\mu$.
This unit vector defines a one-dimensional real space,
which we extend to a one-dimensional complex space.
Further, $\Gamma(y)$ is a rectangle with corners
$-i\epsilon,\bar{y}-i\epsilon,\bar{y}+i\epsilon,i\epsilon$,
with negative orientation, cf.~Fig.(\ref{figf}).
$\epsilon$ is a fixed strictly positive real parameter,
which need not be small.
This being said, Eq.(\ref{cont}) is obvious.

Next, we define the parametrization of $\Gamma(y)$
by four functions $f_i(u,k,y;\rho,\epsilon)$
\begin{eqnarray}
\oint_{\Gamma(y)}dz e^{iz k\cdot \hat{y}} & \equiv &
\sum_{i=1}^4\int_{l_i}^{u_i} du  f_i(u,k,y;\rho,\epsilon)\nonumber\\
f_1(u,k,y;\rho,\epsilon) & := &
e^{ik\cdot \hat{y}(u-i\epsilon)},\;l_1=\rho\bar{y},\;u_1=\bar{y},
\nonumber\\
f_2(u,k,y;\rho,\epsilon) & := & e^{ik\cdot \hat{y}(\bar{y}+ui)},\;l_2=
-\epsilon,\;u_2=\epsilon,\nonumber\\
f_3(u,k,y;\rho,\epsilon) & := & -e^{ik\cdot \hat{y}(u+i\epsilon)},\;l_3=\rho\bar{y},
\;u_3=\bar{y},
\nonumber\\
f_4(u,k,y;\rho,\epsilon) & := & -e^{ik\cdot \hat{y}(\rho\bar{y}+ui)},\;l_4=
-\epsilon,\;u_4=\epsilon,\label{ci}
\end{eqnarray}
where $\rho$ is
a small real positive parameter introduced for convenience.
We will discuss the limit $\rho\to 0$ below.
Occasionally, we will abbreviate $f(u,k,y;\rho,\epsilon)$ as $f(k,y)$.

We now define ten contour integrals $I_i$ which all vanish
by Eq.(\ref{cont}). We exploit the fact that a propagator
$\Delta_F(x_1-x_2)$ provides a
vector $y=x_1-x_2$, which can serve to
define the parallel space which we want to complexify.
By construction, our approach demands that the contour
$\Gamma(y)$ runs in the complexified real
space which is defined by  a propagator which connects
two vertices. The very presence of such a propagator
allows the definition of the one-dimensional
parallel space. So we cannot connect  vertices at
two arbitrarily chosen points $x,y$
by a contour, but have to choose vertices which are connected by a
propagator $\Delta_F(x-y)$.

All our contours $\Gamma$ will have the property that they
consists of two sub-paths $\gamma_{i,1}$ and $\gamma_{i,3}$,
which run parallel to but above ($\Im(z)>0$) or below ($\Im(z)<0$)
the real axis. We call them horizontal
paths. The two other sub-paths $\gamma_{i,2}$ and $\gamma_{i,4}$
run parallel to the imaginary axis through the endpoints of the
parallel paths.

If we want to stress the $\rho$ dependence of
various integrals $I_i$, we write $I_i(\rho)$.
Occasionally, we will also consider the $\rho$ dependence
of individual integrals $I_{i,j}(\rho)$, where the second
subscript refers to the enumeration of $f_j$ in $I_i$.

We group the ten contour integrals in three classes.
The first class provides $I_1,I_5,I_6,I_{10}$:
\begin{eqnarray}
0=I_1 & := &
\int
\frac{d^4k}{k^2}
\frac{d^4r}{r^2}
  \int dX
G(x_a,x_b,x_c,x_d,x_0)\nonumber\\
 & & 
\times\Delta_F^5(x_r-x_l,x_b-x_l,x_m-x_a,x_d-x_r,x_r-x_c)\label{class1s}\\
 & &
\times e^{ik\cdot (x_l-x_m)}
e^{ir\cdot (x_0-x_l)}
\left[
\sum_{i=1}^4\int_{l_i}^{u_i}du f_i(-r,x_m-x_l)\right],\nonumber\\
0=I_5 & := & -\int \frac{d^4k}{k^2}\frac{d^4r}{r^2}  \int dX
G(x_a,x_b,x_c,x_d,x_0)\nonumber\\
 & & 
\times
\Delta_F^5(x_r-x_l,x_b-x_l,x_l-x_a,x_d-x_m,x_r-x_c)\\
 & & \times e^{ik\cdot (x_m-x_r)}e^{ir\cdot (x_0-x_r)}
\left[  \sum_{i=1}^4\int_{l_i}^{u_i}du f_i(-r,x_m-x_r)\right],\nonumber\\
0=I_6 & := & \int \frac{d^4k}{k^2}\frac{d^4r}{r^2}  \int dX
G(x_a,x_b,x_c,x_d,x_0)\nonumber\\
 & & 
\times
\Delta_F^5(x_r-x_l,x_b-x_l,x_l-x_a,x_d-x_r,x_m-x_c)\\
 & & \times e^{ik\cdot (x_r-x_m)}e^{ir\cdot (x_0-x_r)}
\left[  \sum_{i=1}^4\int_{l_i}^{u_i}du f_i(-r,x_m-x_r)\right],\nonumber\\
0=I_{10} & := & -\int \frac{d^4k}{k^2}\frac{d^4r}{r^2}  \int dX
G(x_a,x_b,x_c,x_d,x_0)\nonumber\\
 & & 
\times
\Delta_F^5(x_r-x_l,x_b-x_m,x_l-x_a,x_d-x_r,x_r-x_c)\label{class1e}\\
 & & \times e^{ik\cdot (x_m-x_l)}e^{ir\cdot (x_0-x_l)}
\left[  \sum_{i=1}^4\int_{l_i}^{u_i}du f_i(-r,x_m-x_l)\right],\nonumber
\end{eqnarray}
where $\Delta^5_F(x_r-x_l,x_b-x_l,\ldots,x_r-x_c)$ denotes
the product of five propagators
$\Delta_F(x_r-x_l)\Delta_F(x_b-x_l),\ldots,\Delta_F(x_r-x_c)$.
The signs in front of the definitions of $I_5$ and $I_{10}$ are determined
by the various orientations and will be explained below.

We mentioned already that
the 4TR has a distinguished propagator, which has an endpoint
$x_m$ which couples at four different points in the four terms.
Referring to Fig.(3), we call this point $z$ when it
moves along  one of the four contour integrals above.
Each of the four cases in this figure corresponds to one
of the four terms in the 4TR, in the sense that the contours
reproduce the Feynman integral when the integrand
of the contour for $f_2$ is evaluated at $u=0$.
There, $f_2$ evaluates to $e^{iy\cdot k}$ (for generic $y,k$)
and thus reproduces the propagator $\Delta_F(x_0-x_m)$
upon inverse Fourier transformation.
So, at $u=0$,
the function $f_2$ in $I_1$  reproduces the Feynman graph
${\cal G}_1$, and similarly
$I_{10}$ matches ${\cal G}_2$, $I_6$ matches ${\cal G}_3$
and $I_{5}$ matches ${\cal G}_4$.

Note that when moving this endpoint along, there still remains
a two-point vertex at the point $x_m$. Note further
that our contours move the point $z$
close to one of the endpoints of the propagator $\Delta_F(x_l-x_r)$;
the point $z$ will
actually traverse one of these endpoints
in the limit $\rho\to 0$ at the point $u=0$ in
$f_4$.

The second class provides $I_2,I_4,I_7,I_9$:
\begin{eqnarray}
0=I_2 & := & \int \frac{d^4k}{k^2}\frac{d^4r}{r^2}  \int dX
G(x_a,x_b,x_c,x_d,x_0)\nonumber\\
 & & 
\times\Delta_F^5(x_r-x_l,x_0-x_l,x_b-x_l,
x_d-x_r,x_r-x_c)\label{class2s}\\
 & & \times e^{ir\cdot (x_l-x_a)}
\left[  \sum_{i=1}^4\int_{l_i}^{u_i}du
f_i(r-k,x_m-x_l)\right],\nonumber\\
0=I_4 & := & -\int \frac{d^4k}{k^2}\frac{d^4r}{r^2}  \int dX
G(x_a,x_b,x_c,x_d,x_0)\nonumber\\
 & & 
\times
\Delta_F^5(x_r-x_l,x_0-x_r,x_b-x_l,x_l-x_a,x_r-x_c)\\
 & & \times e^{ir\cdot (x_d-x_r)}
\left[  \sum_{i=1}^4\int_{l_i}^{u_i}du
f_i(k-r,x_m-x_r)
\right],\nonumber\\
0=I_7 & := & \int \frac{d^4k}{k^2}\frac{d^4r}{r^2}  \int dX
G(x_a,x_b,x_c,x_d,x_0)\nonumber\\
 & & 
\times
\Delta_F^5(x_r-x_l,x_0-x_r,x_b-x_l,x_l-x_a,x_d-x_r)\\
 & & \times e^{ir\cdot (x_r-x_c)}
\left[  \sum_{i=1}^4\int_{l_i}^{u_i}du
f_i(r-k,x_m-x_r)
\right],\nonumber\\
0=I_9 & := & -\int \frac{d^4k}{k^2}\frac{d^4r}{r^2}  \int dX
G(x_a,x_b,x_c,x_d,x_0)\nonumber\\
 & & 
\times
\Delta_F^5(x_r-x_l,x_0-x_l,x_l-x_a,x_d-x_r,x_r-x_c)\label{class2e}\\
 & & \times e^{ir\cdot (x_b-x_l)}
\left[  \sum_{i=1}^4\int_{l_i}^{u_i}du
f_i(k-r,x_m-x_l)
\right].\nonumber
\end{eqnarray}
This time, it is the two-point vertex at $x_m$ itself which moves
along a similar contour. It is the sole purpose
of the contours in the second class  to move this point
close to $x_l$ (on the ``left'') or $x_r$ (on the ``right'').
We need it there to make contact with the curves
in the third class. As we move
the whole two-point vertex, the functions $f_i$ will depend
on two momenta in the form $k-r$ or $r-k$. As long as $\rho\not=0$,
the point $x_m$ will not merge with either $x_l$ or $x_r$.
This is in fact a very dangerous point. The moment when $x_m$ collapses
to one of these points, the total number of propagators in our Feynman
diagram is reduced by one, and thus the powercounting can change drastically.
Especially, it might and will in general happen that we generate
subdivergences in the collapse. This is what really makes the 4TR
a {\em four}-term relation. We will see that we can arrange
only four contours in a manner so that the collapse at $x_l$ or $x_r$
always happens twice, with opposite signs.
Hence, two contours stabilize the collapse at $x_l$ and two further ones
at $x_r$.
These cancellations
then allow for the limit $\rho \to 0$ at the end, and generate the
4TR with the promised signs.

The third class defines $I_3,I_8$. It slightly
modifies the $f_i$ and only becomes a proper contour
integral in the limit $\rho\to 0$:
\begin{eqnarray}
0=\lim_{\rho\to 0}I_3 & := &
-\int \frac{d^4k}{k^2}\frac{d^4r}{r^2}
\frac{d^4m}{m^2}
\frac{d^4m_1}{m_1^2}\frac{d^4m_2}{m_2^2}
\int dX
G(x_a,x_b,x_c,x_d,x_0)\nonumber\\
 & & 
\times
\Delta_F(x_b-x_l)\Delta_F(x_r-x_c)\label{class3s}\\
 & & \times e^{ik\cdot (x_r-x_l)}
e^{im_1\cdot (x_l-x_a)}e^{im_2\cdot (x_d-x_r)}
e^{ir\cdot (x_0-x_l)}\nonumber\\
 & & 
\times
\left[  \sum_{i=1}^4\int_{l_i}^{u_i}du
\tilde{f_i}(u,r,m_1,m_2,m,x_l-x_r;\rho,\epsilon)\right],
\nonumber\\
0=\lim_{\rho\to 0}I_8 & := &
\int \frac{d^4k}{k^2}\frac{d^4r}{r^2}
\frac{d^4m}{m^2}  \frac{d^4m_1}{m_1^2}\frac{d^4m_2}{m_2^2}
\int dX
G(x_a,x_b,x_c,x_d,x_0)\nonumber\\
 & & 
\times
\Delta_F(x_l-x_a)\Delta_F(x_d-x_r)\label{class3e}\\
 & & \times e^{ik\cdot (x_r-x_l)}
e^{im_1\cdot (x_b-x_l)}e^{im_2\cdot (x_r-x_c)}
e^{ir\cdot (x_0-x_l)}\nonumber\\
 & & 
\times
\left[  \sum_{i=1}^4\int_{l_i}^{u_i}du
\tilde{f_i}(u,r,m_1,m_2,m,x_l-x_r;\rho,\epsilon)\right],
\nonumber
\end{eqnarray}
where $\tilde{f}$ is defined below.
These two contour integrals once more move the same distinguished
propagator as the contours
in the first class.
In the limit $\rho\to 0$, all dependence on the location of the point
$x_m$ disappears. It is either collapsed to $x_l$ or $x_r$.

Note that this time we explicitly gave five propagators
in Fourier transformed form in Eq.(\ref{class3s},\ref{class3e}).
However, the total number of propagators
that are not summarized in $G(x_a,x_b,x_c,x_d,x_0)$ is still
seven, since two remain
in coordinate form.
This time we use the propagator
$\Delta_F(x_l-x_r)$, present in all four terms, to define
the parallel space and to connect
the ``left side'' ($x_a,x_b$) with the ``right side'' $(x_c, x_d)$.
For $I_3$, this happens while $x_m$ approaches the collapse
either along the propagators $\Delta_F(x_l-x_a)$ or $\Delta_F(x_d-x_r)$,
while for $I_8$, $x_m$ runs either along $\Delta_F(x_b-x_l)$ or
$\Delta_F(x_c-x_r)$. Our orientations are such that all vertical paths
which coincide in the limit $\rho\to 0$ have opposite
orientations.

We note that in the limit $\rho\to 0$
\begin{eqnarray}
\tilde{f_i}(u,r,m_1,m_2,m,x_l-x_r;\rho,\epsilon)
=f_i(u,r,x_l-x_r;\rho,\epsilon),\;\;\forall m_1,
m_2,m,
\end{eqnarray}
according to the following definition of
$\tilde{f_i}(u,r,m_1,m_2,m,x_l-x_r;
\rho,\epsilon)$
\begin{eqnarray}
0 & < & \rho  <<  1,\nonumber\\
e_1 & := & x_m-x_l,\; \mbox{for}\;I_3,\;\;x_l-x_m,\; \mbox{for}\;I_8,
\nonumber\\
e_2 & := & x_r-x_m,\; \mbox{for}\;I_3,\;\;x_m-x_r,\; \mbox{for}\;I_8,
\nonumber\\
\tilde{f_2}(u,r,m_1,m_2,m,x_l-x_r;\rho,\epsilon) & := &
f_2(u,r,x_l-x_r;0,\epsilon)e^{i\rho(m_1+m)\cdot e_1},\label{ftil}\\
\tilde{f_4}(u,r,m_1,m_2,m,x_l-x_r;\rho,\epsilon) & := &
f_4(u,r,x_l-x_r;0,\epsilon)e^{i\rho(m_2+m)\cdot e_2},\nonumber\\
\tilde{f_i}(u,r,m_1,m_2,m,x_l-x_r;\rho,\epsilon) & := &
f_i(u,r,x_l-x_r;0,\epsilon)
\;\;\forall i\in\;\{1,3\}.\nonumber
\end{eqnarray}
In the limit $\rho\to 0$ the $m$ integration
in $I_3,I_8$ as well as the $k$ integration
in $I_{2,4},I_{4,4},I_{7,4}$ and $I_{9,4}$ decouples.
It is easy to see that such integrations are independent of
the orientation of the vector $e=e_{1,2}$ defining the parallel space,
due to rotational invariance of the measure $\int d^4m$:
\begin{equation}
\int d^4m \frac{e^{i\rho m\cdot e}}{m^2}
\sim\int dm_\parallel \sqrt{m_\parallel^2}[
e^{i\rho \overline{e}m_\parallel}
+e^{-i\rho \overline{e}m_\parallel}],\;\forall e,\label{dec}
\end{equation}
We do not worry at this point that this
integral is ill-defined. We use a cut-off $\lambda$ and will
later see that all these contributions vanish.

Note further that with these definitions the horizontal paths
in class three cancel out:
\begin{equation}
I_{3,1}+I_{8,1}=I_{3,3}+I_{8,3}=0,\label{hor3}
\end{equation}
by construction. For these horizontal paths, the $f_i$´s
remain $\rho$-independent, and thus Eq.(\ref{hor3})
follows from using Eq.(\ref{fourier}) backwards
for the $m_1,m_2$ integrations.

We already stressed that the integrals $I_{1,2},I_{5,2},I_{6,2},I_{10,2}$
contain the Feynman integrals
${\cal G}_i$ at the point $u=0$. They are in fact ill-defined
at this point if we let the cut-off in the final
$\int d^4k$ integration go to infinity.
It is our main purpose  to show that the only
remaining contributions which diverge for large cut-off come
from these $I_{i,2}$. Thus, what has to be shown is that neither
the horizontal paths diverge at the end, nor the vertical paths
apart from the ones in Eq.(\ref{match}).
Not unexpectedly, the horizontal paths will become finite
when one considers the paths above and below the real axis together,
and the vertical paths cancel in appropriate pairs (or quartets)
due to their
opposite orientations.

To this end, we show
that the following four propositions hold.
\begin{prop}
For $i\in \{1,2,4,5,6,7,9,10\}$ we have
\begin{equation}
<I_{i,1}(\rho)+I_{i,3}(\rho)>=0,\;\;\forall \rho>0.
\end{equation}
\end{prop}
Proof:
For our horizontal paths, the endpoints of the $u$-integrations
depend on  coordinates $x_l,x_r,x_m$. We thus cannot interchange
the $u$-integration with these integrations. Accordingly, we will
do the $u$-integration first.

But we can integrate last the momentum which appears
in the $f$-function. Let us call this momentum
$k$. For this momentum integration,
we will use the measure $\int d^4k$  separated in parallel and orthogonal
space integrations.

First, integrating $u$ delivers
\begin{equation}
\int_{\rho \overline{y}}^{\overline{y}}du e^{iu k_\parallel}
=\frac{e^{iy\cdot k}-e^{iy\cdot \rho k}}{ik_\parallel},
\label{uint}
\end{equation}
where $k_\parallel$ is the component of the four-momentum
$k$ in the direction of the four-vector $y$, which
is one of the differences of coordinates as they appear
in the functions $f_1,f_3$ in the integrals in the proposition.
After the $u$ integration, we will do all remaining integrals
but the $k$ integration. Inspection of Eqs.(\ref{uint})
and (\ref{class1s}-\ref{class1e},\ref{class3s},\ref{class3e})
shows that the integrand
is a proper Feynman integral concerning these integrations,
with $k$ or $\rho k$ acting as an exterior momentum for these
integrations.

Doing all integrations but the $k$ integration,
we know from powercounting that
there remains a factor $d^4k/k^4$,
expressing the overall logarithmic divergence.
Thus, the results of all but the last integration
is to multiply Eq.(\ref{uint}) by $c(\rho)d^4k/k^4$,
where $c(\rho)$ is a pure number obtained from finite integrations
over coordinates and momenta.

For the practitioner of computational field theory, the calculation
of the penultimate integration is the end of the story. The finite
value so obtained ($c(\rho)$ in our case) determines the coefficient
of the divergence in the final momentum integration.
The divergence would be provided by a behaviour $\sim \log(\lambda)$
if we use a cut-off $\lambda$ in the final integration, for a
logarithmic divergent Feynman integral. What for us remains to do is to
check if the final momentum integration provides a divergence
along the horizontal paths.

From the $u$-integration of Eq.(\ref{uint}) and the overall factor of
$c(\rho)d^4k/k^4$, we find that the paired
horizontal paths deliver
\begin{eqnarray}
 & & c(\rho)\int_{-\lambda}^\lambda dk_\parallel \int d^{3}k_\perp
\left[
\frac{e^{-\epsilon k_\parallel}-e^{\epsilon k_\parallel}}{k_\parallel}
\right]
\frac{1}{[k^2]^2}
\sim c(\rho)
\int_{-\lambda}^{\lambda}
dk_\parallel
\left[
\frac{e^{-\epsilon k_\parallel}-e^{\epsilon k_\parallel}}{k_\parallel}
\right]
\frac{1}{\sqrt{k_\parallel^2}},\label{cr}
\end{eqnarray}
where $k^2=k_\parallel^2+k_\perp^2$,
and the exponentials resulted from
the appearance of $\epsilon$ in Eq.(\ref{ci}).\footnote{We
work in Euclidean
space for convenience. It is possible to work in
Minkowski space as well. One merely has to show that
for $k_\parallel$ timelike or spacelike similar
expressions result.}
We stress that in the final $k_\parallel$ integration
we are only interested in the (UV-) behaviour for large $k_\parallel$.
The behaviour near $k_\parallel=0$ is of no interest for us.
Accordingly, we nullified all external parameters.
This has not changed the UV behaviour at all, but for the
behaviour near $k_\parallel=0$ we should have in mind that
the IR behaviour is in fact dependent on the value of external
parameters like masses and momenta.\footnote{In
 dimensional regularization, we would simply keep a scale
in the final momentum integration.} The pole
at $k_\parallel=0$ is in fact fictitious
and absent if we keep a mass or momentum
$\not=0$. Equivalently,
from now on it is implicitly understood that we use a
 small IR cut-off in the final $k_\parallel$ integration.

It may appear odd that the final momentum integration
involves a non-covariant separation between
$k_\parallel$ and $k_\perp$.
To see how this may arise, consider the following integral
\begin{eqnarray}
 & & \int d^4x \int d^4k  \int du\; F(k^2,x^2,k\cdot x,u)\nonumber\\
 & =  &  \int d\Omega_{\hat{x}}
\int d^4k  \int du\int d\overline{x}\;\overline{x}^3
\; F(k^2,x^2,k\cdot x,u)\label{F2}\\
 & = & \int d\Omega_{\hat{x}}
\int d^4k  \int du \;
G(k^2,k\cdot \hat{x},u)\label{G1}\\
 & = & 2\pi^2
\int d^4k  \int du \;
G(k^2,k_\parallel,u)\label{G2}.
\end{eqnarray}
After all but two of the integrations have been
performed, one obtains a covariant integral over a  coordinate
$x$ and momentum $k$ and a contour variable $u$,
whose integrand $F$ can depend only on the three scalar products
and $u$. In Eq.(\ref{F2}) we perform first the integral over
$\overline{x}:=\sqrt{x^2}$.
In Eq.(\ref{G1}) the result of the $\overline{x}$
integration must yield a function $G$ that depends only on
$k^2$, the component of $k$ in the direction of the
unit vector $\hat{x}$ and $u$.
Since the $d^4k$ integration can leave no dependence on the
direction of $\hat{x}$, in Eq.(\ref{G2})
we may replace the angular integration
by the value, $2\pi^2$, of the surface of the unit-sphere
in four dimensions,
and also replace $k\cdot\hat{x}$ by $k_\parallel$
which is a component of $k$ along {\em any}
fixed axis.

By construction, $c(\rho)$ is a finite quantity for all
$\rho\not=0$. What remains to prove the theorem is to show
that the integral in Eq.(\ref{cr}) remains finite for large $\lambda$.

\begin{figure}
\epsfysize=4cm
\epsfbox{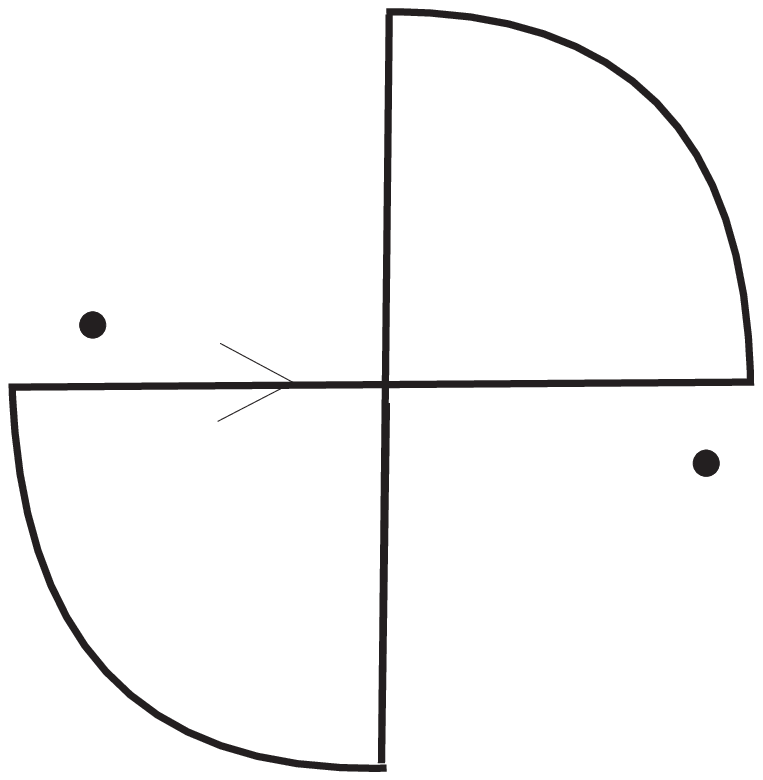}
\caption{\label{figcont}The contour rotation.
The two dots indicate the endpoints
of the cuts from
$(k_\parallel^2+i \sigma)^{1/2}$.
}
\end{figure}

We proceed with the  $k_\parallel$ integration.
Due to the factor
$[e^{- \epsilon k_\parallel}-e^{ \epsilon
k_\parallel}]/k_\parallel$ this
is not a standard loop integration. We achieve the $k_\parallel$
integration, which is along the real axis, by analytic continuation
to the imaginary axis in complex $k_\parallel$ space.
The integration
$\int_{-\lambda}^\lambda dk_\parallel$ becomes a closed contour by adding
an arc with radius $\lambda$ from $+\lambda$ to $+i\lambda$
(counterclockwise!),
integrating along the imaginary axis from $+i\lambda$ to $ -i\lambda$
and another arc from $-i\lambda$ to $-\lambda$
(clockwise).
We read $\sqrt{k_\parallel^2}$ above as
$\sqrt{k_\parallel^2+i\sigma}$, $0<\sigma<<1$, as this does not change
the value of the integral. Then the integrand remains single-valued
along and inside the contour in Fig.(\ref{figcont}).

We conclude that the integration from $-\lambda$ to
$\lambda$ is given by the three terms above.
We show next that the two arcs add to a finite contribution.
With $z=\lambda e^{i\phi}$ we obtain
\begin{equation}
\frac{i}{\lambda}\int_0^{\pi/2} d\phi
\left[e^{\lambda e^{i\phi}}e^{-i\phi}-e^{-\lambda e^{i\phi}}e^{-i\phi}
\right]
-
\frac{i}{\lambda}\int_{\pi}^{3\pi/2} d\phi
\left[e^{\lambda e^{i\phi}}e^{-i\phi}-e^{-\lambda e^{i\phi}}e^{-i\phi}
\right],
\end{equation}
which adds to zero upon setting $\phi^\prime=\phi-\pi$
in the second integration.

One readily verifies that the integration along the imaginary
axis is finite.
It provides an expression
\begin{eqnarray}
i c(\rho)
\int_{0}^{\lambda} dk_\parallel
\left[
\frac{e^{-i \epsilon k_\parallel}-e^{i \epsilon k_\parallel}}{k_\parallel}
\right]
\frac{1}{\sqrt{k_\parallel^2}}<\infty.
\end{eqnarray}
This completes the proof.\hfill$\Box$\\
This proposition guarantees that for $\rho>0$ the horizontal paths
in the proposition are well defined when our momentum
integrations tend to infinity. Further, it follows that
$<I_{j,2}+I_{j,4}>=0$, $\forall j\in \{2,4,7,9\}$,
$\forall \rho>0$.

Next, we can prove the following proposition.
\begin{prop}
We have
\begin{equation}
<I_{1,4}+I_{3,2}>=<I_{5,4}+I_{3,4}>=<I_{6,4}+I_{8,4}>
=<I_{10,4}+I_{8,2}>=0,\;\;\forall \rho > 0 .
\end{equation}
\end{prop}
Proof: The proposition compares integrals
of the first class with integrals of the third class in pairs.
In these pairs, the elements of the first class are distinguished
from elements of the third class by two facts: they use
different propagators to define the parallel space,
and the point $x_m$ is at different locations.

But the point $x_m$ is only a two-point vertex.
From our experience with integrals in the second class we
know that we can transport it freely as long as it does not collapse
to either $x_l$ or $x_r$.

Hence, for all integrals $I_{i,4}$ in the first class
we use contour integrals as provided by the second class to
transport the point $x_m$. We move it  so that it is in the
same position as in the integrals of the third class
in the proposition.

Next, we assume that we carried out all
$\int dX$ integrations.
The remaining
integrations concern $\int du$ and momentum
integrations.\footnote{Here, we use that
the $u$ integration is decoupled from the integration
of the coordinates for vertical paths.
We further exchange orders of integrations at will,
which is legitimate for any $\lambda<\infty$
and $\rho>0$, where our Feynman integrands
are fairly well-defined analytic functions.}
We now proceed by the following
argument. A diagram
which is free of subdivergences can be opened at any line,
its value is independent from the point where we cut the diagram.
The finite value assigned to the diagram after the
penultimate integration  appears as the coefficient of
divergence in the final loop integration and is uniquely
determined  after the penultimate
integration is carried out. It is only the final
integration which diverges (in subdivergence free diagrams).

Now, we cut the diagram at the propagator which involves
the factor $e^{-u\epsilon r_\parallel}$. This has the advantage
that the remaining expression is $u$-independent.
All $u$ dependence affects only the final loop
integration. Such a propagator always exist,
by inspection of Eqs.(\ref{class1s}-\ref{class1e})
and of Eqs.(\ref{class2s}-\ref{class2e}).
As a consequence, only the final momentum integration depends
on the variable $u$, and will be of the general form
\begin{equation}
\int_{-1}^1 du \int d^4r \frac{c e^{-\epsilon u r_\parallel}}{[r^2]^2},
\end{equation}
as our overall degree of divergence was logarithmic.
Here, $c$ is the uniquely determined $u$-independent
value of the diagram after all
momentum integrations but the final one are carried out.
The $u$ integration runs in this parametrization
from $\int_{-1}^1 du$ for vertical paths from below to above
the real axis, and vice versa for vertical paths with the
opposite orientation.

But the vertical contours
which are paired in the proposition have opposite orientation,
but otherwise their integrands must agree after
the penultimate integration. This proves
the proposition.\footnote{A different proof may be obtained
by isolating the singularity at $u=0$ in
appropriate functions $f_2$ or $f_4$. At these points,
the paired integrands match and cancel out. Isolating
the singularities is most easily obtained by replacing the complex plane
by ${\bf R}^2$, with $u\to iu$,
which effectively introduces rapidly oscillating
integrands whenever $u\not=0$ in vertical paths.}\hfill$\Box$\\

\begin{figure}
\epsfysize=4cm
\epsfbox{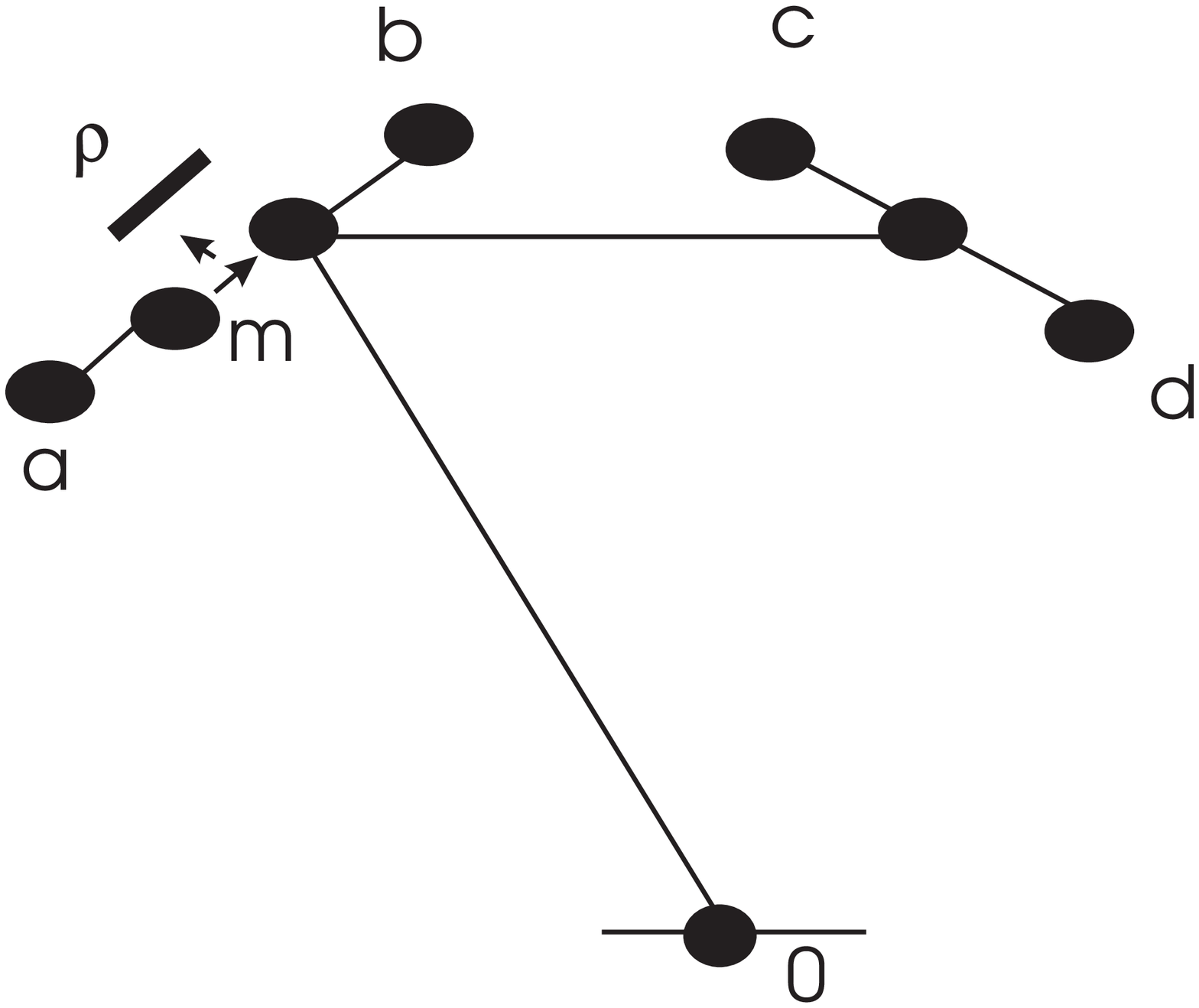}
\caption{\label{figsub}The appearance of a subdivergence at $\rho=0$.}
\end{figure}
\begin{figure}
\epsfxsize=47mm \epsfbox{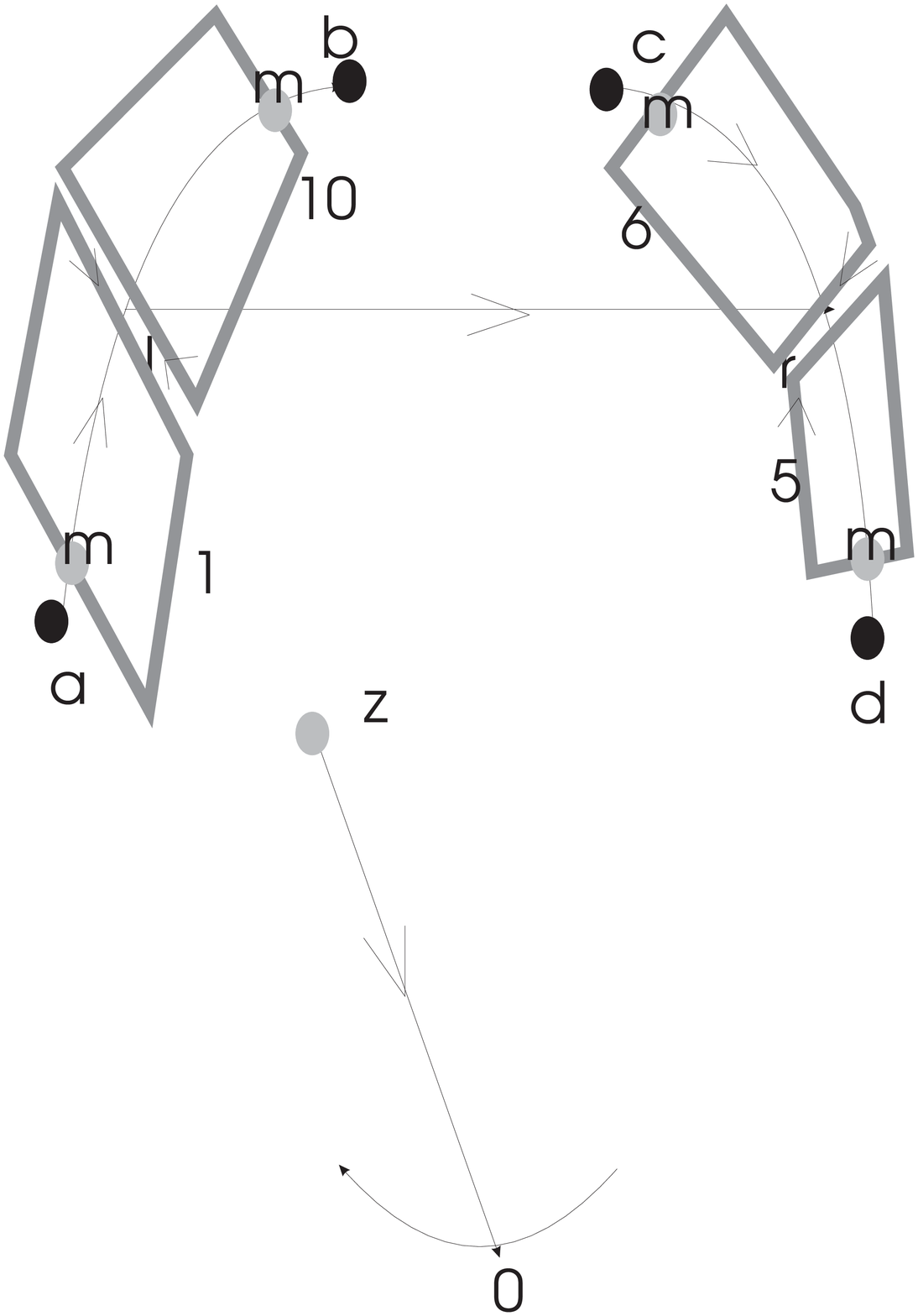}
\caption{\label{figor1}The orientations for the first class
$I_1$, $I_5$, $I_6$, $I_{10}$.
Orientations are chosen in accordance with Eq.(\ref{ci})
and Eqs.(\ref{class1s}-\ref{class3e})
such that
the various propositions hold. For class
1 all contours are oriented clockwise;
the grey dots denote the point $x_m$ in each contour.
The point $z$ moves along the contours. At $z=x_m$ we recover
a genuine Feynman graph. The contours will cross the points
$x_l$ or $x_r$ if $\rho=0$.}
\end{figure}
\begin{figure}
\epsfxsize=47mm \epsfbox{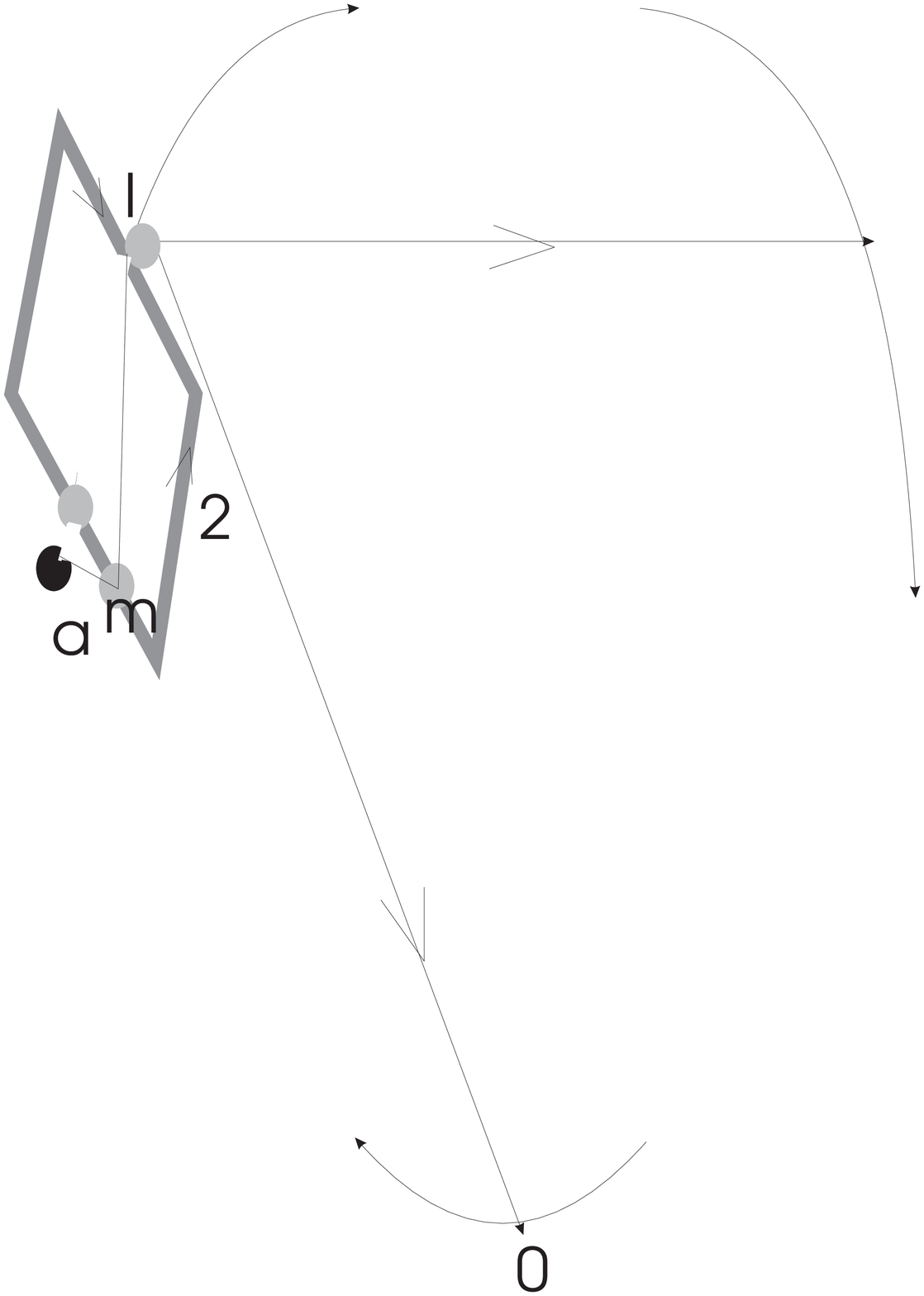}
\caption{\label{figor2}$I_2$ as an example for the second class.
$I_4$, $I_7$, $I_9$ are similar and located in the same places
as their partners from the first class in Fig.(\ref{figor1}).
By comparison with the previous figure,
we realize that it is now the former $x_m$ which moves.
For $\rho=0$, the contour traverses $x_l$.}
\end{figure}
\begin{figure}
\epsfxsize=47mm \epsfbox{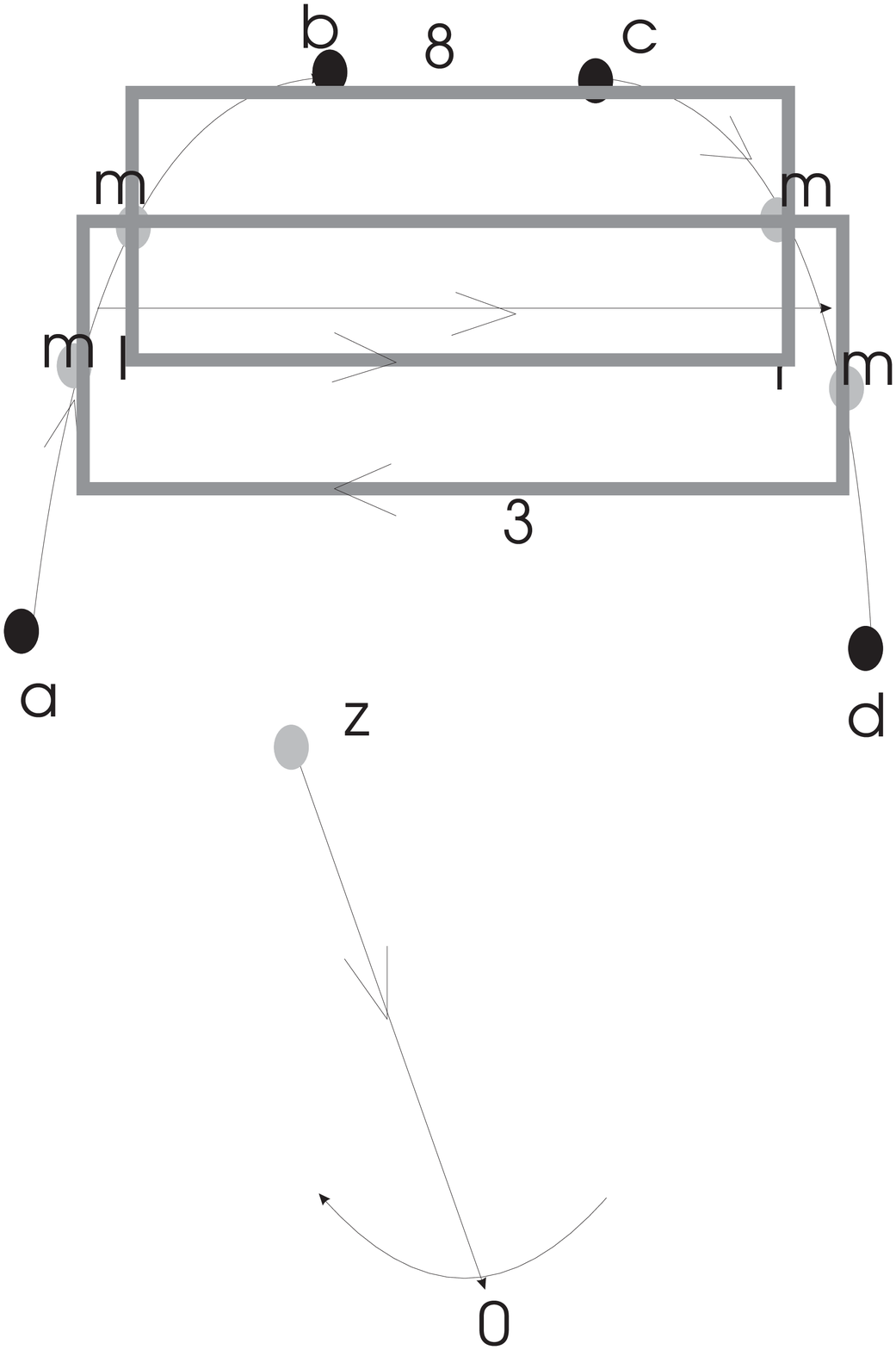}
\caption{\label{figor3}The two grey rectangles indicate the two contours
$I_3,I_8$. They coincide and cancel at $\rho=0$.
The point $x_m$ moves from the left to the right
as long as $\rho\not=0$, but disappears at $\rho=0$.}
\end{figure}

Our next proposition involves the limit $\rho\to 0$.
The main obstacle in this limit is that subdivergences may appear.
In this limit, we effectively decoupled a propagator.
The weight of this propagator is thus missing in the powercounting
for the remainder. This, in general, will generate a subdivergence.
We will see that with our choice of orientations the relevant
combinations of contours still allow the limit $\rho\to 0$.
In Fig.(\ref{figsub}) we demonstrate the presence of the subdivergence
at $\rho=0$.
Figs.(\ref{figor1},\ref{figor2},\ref{figor3})
demonstrate the various orientations of the contours.

We collect all combinations of contributions which have to cancel
for $\rho\to 0$ in a proposition.
\begin{prop}
In the limit of $\rho\to 0$ we have
\begin{eqnarray}
<I_{1,1}+I_{1,3}+I_{10,1}+I_{10,3}> & = & 0\;\;\mbox{(Ia)},\nonumber\\
<I_{5,1}+I_{5,3}+I_{6,1}+I_{6,3}> & = & 0\;\;\mbox{(Ib)},\nonumber\\
<I_{2,1}+I_{2,3}+I_{9,1}+I_{9,3}> & = & 0\;\;\mbox{(IIa)},\nonumber\\
<I_{7,1}+I_{7,3}+I_{4,1}+I_{4,3}> & = & 0\;\;\mbox{(IIb)},\nonumber\\
<I_{1,4}+I_{3,2}+I_{10,4}+I_{8,2}> & = & 0\;\;\mbox{(IIIa)},\nonumber\\
<I_{5,4}+I_{3,4}+I_{6,4}+I_{8,4}> & = & 0\;\;\mbox{(IIIb)}\nonumber.
\end{eqnarray}
\end{prop}
Proof:
Following our general philosophy, we assume that
all momentum integrations are cut-off by an
appropriate $\lambda$. Then, all analytic expressions
are well-defined $\forall \rho$ and have a finite
limit when $\rho$ tends to zero.
We thus only have to check that the above proposition
combines integrals such that they approach the same limit
when $\rho\to 0$, but with
opposite orientations, so that they identically cancel,
cf.~Figs.(\ref{figor1},\ref{figor2},\ref{figor3}).
We then let $\lambda\to\infty$ afterwards.

For (Ia), we note that $I_{1,1}+I_{1,3}$ is well defined for
all $\epsilon$, as well as $I_{10,1}+I_{10,3}$, and for all $\rho\not=0$,
by Prop.(1). At $\rho=0$, the two pairs approach the same limit,
by inspecting Eqs.(\ref{class1s}-\ref{class1e}).
But at $\rho=0$, the
integrands of the two pairs equal and the
contours are oriented in the opposite manner. Thus they cancel
identically. Similar arguments hold for case (Ib).

For (IIa,IIb) again a similar argument holds,
inspecting Eqs.(\ref{class2s}-\ref{class2e}).

For cases (IIIa) and (IIIb) once more, the given combinations
of integrals exist for any $\rho>0$ and cancel out at $\rho=0$.
To see this, one simply considers Eqs.(\ref{class1s}-\ref{class1e})
and (\ref{class3s},\ref{class3e})
and checks the definitions of $\tilde{f_i}$ in Eqs.(\ref{ftil})
and various
orientations in Figs.(\ref{figor1},\ref{figor3}).
In fact, $ \tilde{f}$ was defined such that
its vertical paths $f_4$ in the second class agree with the
oppositely oriented vertical paths in the third class.\hfill$\Box$

Summarizing our considerations, we recover Eq.(\ref{sum})
\begin{eqnarray}
0 & = & <\sum_{i=1}^{10}I_i\nonumber>\\
 & = &  <I_{1,2}+I_{5,2}+I_{6,2}+I_{10,2}>.
\end{eqnarray}
To prove the theorem, it remains to show that these four terms
deliver the overall divergences of ${\cal G}_i$, with the
correct signs.

First of all, we check that the remaining contributions
give the integrals we want.
With $\Lambda:=\int_1^\lambda dt\; e^{\epsilon t}/t^2$, we claim
that the following proposition holds.
\begin{prop}
\begin{equation}
\lim_{\lambda\to\infty}I_{i,2}/\Lambda\;=:\;<I_{i,2}>
\;\sim\;<{\cal G}_r>\;=:\;\lim_{\lambda\to \infty}{\cal G}_r/\log{(\lambda)}
<\infty,
\end{equation}
where the $i$ and $r$ indices match as in Eq.(\ref{match}),
and where our propositions demand that we use the same $\epsilon$
for all $i$.
\end{prop}
Proof: First note that
the four $I_{i,2}$ are $\rho$-independent.
They all appear at the endpoints $u=y$ in vertical paths $f_2$.

We once more cut at the $r$-integration,
cf.~Eq.(\ref{class1s}-\ref{class1e}),
again exploiting that all $u$ dependence is in the final
integration.
It is of the form
\begin{eqnarray}
\lim_{\lambda\to\infty}
I_{i,2} & \sim &
\lim_{\lambda\to\infty}
\int_{-\lambda}^\lambda dr_\parallel \int d^{3}r_\perp
\int_{-\epsilon}^\epsilon du\frac{ce^{-ur_\parallel}}{
[r_\parallel^2+r_\perp^2]^2}
\nonumber\\
 & \sim & \lim_{\lambda\to\infty}
\int_{-\lambda}^\lambda dr_\parallel
\frac{c[e^{\epsilon r_\parallel}
-e^{-\epsilon r_\parallel}]}{r_\parallel\sqrt{r_\parallel^2}}
\sim \lim_{\lambda\to\infty}
\int_1^\lambda dr_\parallel \frac{ce^{
\epsilon r_\parallel}}{r_\parallel^2}=c\lim_{\lambda\to\infty}\Lambda,
\label{lam1}\\
{\cal G}_r & \sim &
\lim_{\lambda\to\infty}\int_1^\lambda dr_\parallel
\frac{c}{r_\parallel}\sim
c\log{(\lambda)},\label{lam2}
\end{eqnarray}
where $c$ is the actual $\epsilon$-independent coefficient of the
overall divergence we are interested in, and which appears after the
penultimate momentum integration is carried out.
We restricted ourselves to terms which diverge for large cut-off
$\lambda$. This justified that we abandoned terms involving
$e^{-\epsilon r_\parallel}$, and that we deliberately
cut off the final
$r_\parallel$ integration at the lower boundary at 1.
Eqs.(\ref{lam1})
and (\ref{lam2}) show that the
vanishing of $<I_{i,2}>$ is equivalent to the vanishing of
the appropriate matched $<{\cal G}_r>$.
Further, with the orientations of the various
contours chosen as demanded by the previous propositions,
the signs are as in Eq.(\ref{match}).\hfill$\Box$

The attentive reader may have asked her/himself
whether the arguments of Prop.(1) could be extended
to prove that each term in the four-term relation
is separately finite, thereby making Prop.(4) trivial.
As an aside, we here indicate the analytical
obstruction that fortunately prevents such a
trivialization.

We have to integrate $u$ from $-\epsilon$ to $\epsilon$.
At $u=0$ we encounter the proper Feynman integral which will
not allow for the limit $\lambda\to\infty$.
It is easy to see that for $u\not=0$ we could analytic continue
 $r_\parallel$ to the imaginary axis,
so that these contributions
($\mid u\mid \in [\eta,\epsilon]$, $0<\eta<<\epsilon$)
remain finite for large cut-off, as we get rapidly oscillating
integrands for large final loop momentum.

Let us now investigate the behaviour near $u=0$.
We once more insert a small imaginary part $i\sigma$ in the
propagators, which will not alter the value of the integral
along the real axis, but guarantees the absence of poles (or cuts)
in the analytic continuation.
In terms of the rescaled four-momentum $p=\eta r$
we investigate the behaviour as $\eta\to 0$
of an expression
of the form
\begin{eqnarray}
 & & \int_{-\lambda}^\lambda dr_\parallel \int_0^\infty
r_\perp^{2}d\!r_\perp
\int_{-\eta}^\eta du \frac{e^{-ur_\parallel}}{
[r_\parallel^2+r_\perp^2+i\sigma]^2}
\nonumber\\
 & = &
\int_{-\lambda}^\lambda dr_\parallel \int_0^\infty r_\perp^{2}d\!r_\perp
 \frac{e^{\eta r_\parallel}-e^{-\eta r_\parallel}}{r_\parallel
[r_\parallel^2+r_\perp^2+i\sigma]^2}
\nonumber\\
 & = &
\eta \int_{-\lambda\eta}^{\lambda\eta}
dp_\parallel \int_0^\infty p_\perp^{2}d\!p_\perp
 \frac{e^{p_\parallel}-e^{-p_\parallel}}{p_\parallel
[p_\parallel^2+p_\perp^2+i\eta^2\sigma]^2
}.\label{pinch}
\end{eqnarray}
The last line above show that
if we want to continue this so that we go from
$e^{p_\parallel}$ to the oscillatory $e^{ip_\parallel}$
we confront a pinch singularity at $\eta=0$.
If we rotated the contour to imaginary $p_\perp$
we would pick up principal values at $(ip_\parallel)^2=p_\perp^2$.

Having concluded the aside, we summarize the derivation of the
four-term relation Eq.(\ref{4tr}) as follows.
\begin{enumerate}
\item
Eq.(\ref{sum}) used contour integration to show that the sum
of 40 integrals necessarily vanishes.
\item
Of these 40, only 4 remain after applying the conclusions of Props.(1-3).
\item
Prop.(1) eliminates the 16 horizontal contributions
of the first two classes.
The 4 horizontal contributions from the third class
cancel each other in Eq.(\ref{hor3}), by definition.
\item
Prop.(2) guarantees that the vertical contributions
of the third class cancel the vertical contributions
$I_{i,4}$, $i\in\{1,5,6,10\}$ of the first class.
\item
Prop.(3) guarantees that, even in the limit $\rho\to 0$,
6 of the 12 cancellations of Props.(1,2) remain valid.
These 6 are sufficient to derive the 4TR.
\item
Prop.(4) identifies the divergences of the 4 remaining integrals
with the 4 counterterms.
\item
The remarks following Prop.(4) remind us of the nontriviality
of Eq.(\ref{sum}), which asserts the finiteness of a very specific combination
of 4 divergent integrals.
\end{enumerate}

Now let us explain our provisos $iii)$ and $iv)$ in
the theorem. These constraints are to be considered as sufficient
conditions. We will try to formulate necessary conditions
in future work.

For the proof of Prop.(3) we had to show that
in the limit $\rho\to 0$ pairs or quartets of
integrals $I_{i,j}$ cancel each other. In this limit, there is
always a propagator which decouples, cf.~Fig(\ref{figsub}).
Implicitly, we assumed so far that this is a scalar propagator.
It provides for any finite $\lambda$ a constant factor,
multiplying the remaining
expression, similar to Eq.(\ref{dec}).

This decoupling was the reason for the possible
existence of a subdivergence.  Now,
this decoupled propagator appears actually at different
places. It appears between the vertex at $x_l$,
and vertices at $x_a$ or $x_b$ in the cases $I_2,I_{3,2}$ (for
$x_a$) or
$I_9,I_{8,2}$ (for $x_b$);
or it appears between the vertex at
$x_r$ and vertices at $x_c$ or $x_d$ in the cases
$I_4,I_{3,4}$ (for $x_d$) or $I_7,I_{8,4}$ (for $x_c$).

It may happen that this propagator $\Delta_F(s)$
is matrix valued, as it
might carry spin-, Lorentz- and/or other indices,
while the vertices $V_l$ and $V_r$ at $x_l$ or $x_r$ might carry indices
as well.

For the contributions to cancel out as demanded in the proof of
proposition (3) we need that
\begin{equation}
[V_i,\Delta_F(s)]=0,\;i\in\{l,r\},\;\forall s^2>0.
\end{equation}
This explains condition $iii)$ in the theorem.
An example in case is when $\Delta_F$ is fermionic, and
the $V_i$ provide vector couplings. We then have that
$[\kslash,\gamma_\mu]\not= 0$, and thus expect the
theorem to fail, due to contributions from UV subdivergences
at $\rho=0$. For example, $I_{j,2}$, $j\in\{3,8\}$ confronts us with
\begin{equation}
\int d^4k \frac{\kslash e^{-u\epsilon k_\parallel}}{k^4}  =
A \hat{\sslash}
\Rightarrow A\sim 2\int_0^\lambda dk_\parallel \left(
e^{-u\epsilon k_\parallel}
-e^{u\epsilon k_\parallel}\right),
\end{equation}
where $\hat{s}$ defines the exterior axis which defines
$k_\parallel$,
so that we obtain as the quantity measuring the failure of the 4TR
\begin{equation}
[\sslash,\gamma_\mu]A.
\end{equation}
Integrating $u$, the remaining expression for $A$ is finite in the limit
$\rho\to 0$, which entails $s^2\to0$.
It thus spoils Prop.(3), as $A$ is a coefficient of logarithmic
divergence.
A more detailed analysis of the general role
of subdivergences will be given in future work.

For condition $iv)$, we note that the presence of spin structures
as
\begin{equation}
\frac{\kslash\gamma_\mu \kslash}{k^4}=
-\frac{\gamma_\mu}{k^2}+2\frac{k_\mu \kslash}{k^4}.\label{tens}
\end{equation}
may make the theorem to fail. In such a case,
one notes that in the limit $s^2\to 0$,
the demanded cancellations  appear in the form $e^{ik\cdot s}-1$,
where now $k$ refers to the final momentum integration
for the subdivergence.
If these factors are accompanied by scalar integrals,
all our arguments go through.  If they are accompanied by
tensors of first degree (which are odd
in $k$), we are safe as this effectively
projects to the odd part of
$e^{ik\cdot s}$, which vanishes at $s^2=0$.

But for tensors of second degree, we are in trouble.
So, while the first term on the rhs of Eq.(\ref{tens})
behaves effectively as a scalar particle
the second term is problematic.
In $e^{ik\cdot s}-1$, the term with the $\exp$-function
allows for a
form-factor expansion
\begin{equation}
\sim A g_{\mu_\nu}+B\hat{s}_\mu\hat{s}_\nu,
\end{equation}
due to the presence
of $\hat{s}$ in $e^{ik\cdot s}$,
which defines a parallel space in the
subdivergent diagram.
The second term without the $\exp$-function only allows for
\begin{equation}
\sim A g_{\mu_\nu}.
\end{equation}
Especially such a situation occurs when a propagator
$x_m-x_l$ (or $x_m-x_r$) is part of a subdivergence only for one of
the two cases which appear paired in proposition (3,(IIIa,b)).
An example is when $I_{3,2}$ provides a subdivergence at
$s^2\not=0$, while $I_{8,2}$ does only so
at $s^2=0$.

This asymmetry provides us
finally with an extra contribution which diverges with $\lambda$,
which can be shown to have the form of
a finite part of a two-loop integral of master topology
\cite{BK4}. A more detailed analysis will be given elsewhere.

\section{A curious case study}
If one wants to find explicit examples for the 4TR in action,
one quickly realizes that there are not many at low loop numbers
\cite{BK4}; for graphs free of subdivergences.
Either the examples at hand are trivial due to symmetries in the graph,
or are very hard to calculate.

In such circumstances, one might wonder if it is possible
to enlarge the number of available examples by
allowing for nonrenormalizable couplings in the universal
function $G(x_a,x_b,x_c,x_d,x_0)$, as this function will not
directly affect the 4TR. In this section, we report on one such example,
where the 4TR is bound to fail as the graphs considered violate one
of our provisos. But we see that in this special case there is a way out,
and the techniques used in this paper extend to manage this case
and predict the results we are interested in. In the course of our
consideration, we will learn that the problem we consider actually is
directly sensitive to the before-mentioned UV-divergence which appears
when one of the propagators decouples. The problem discussed here
is calculated in \cite{BK4}, where the reader will find details of
the calculation and a confirmation of the relation proposed here.

So we start with an
interesting case where our 4TR will fail in its unmodified
form. As envisaged, it is given when one allows for non-renormalizable
couplings in the universal function $G(x_a,\ldots,x_0)$, which
violates proviso $v)$.

\begin{figure}
\epsfysize=4cm\epsfbox{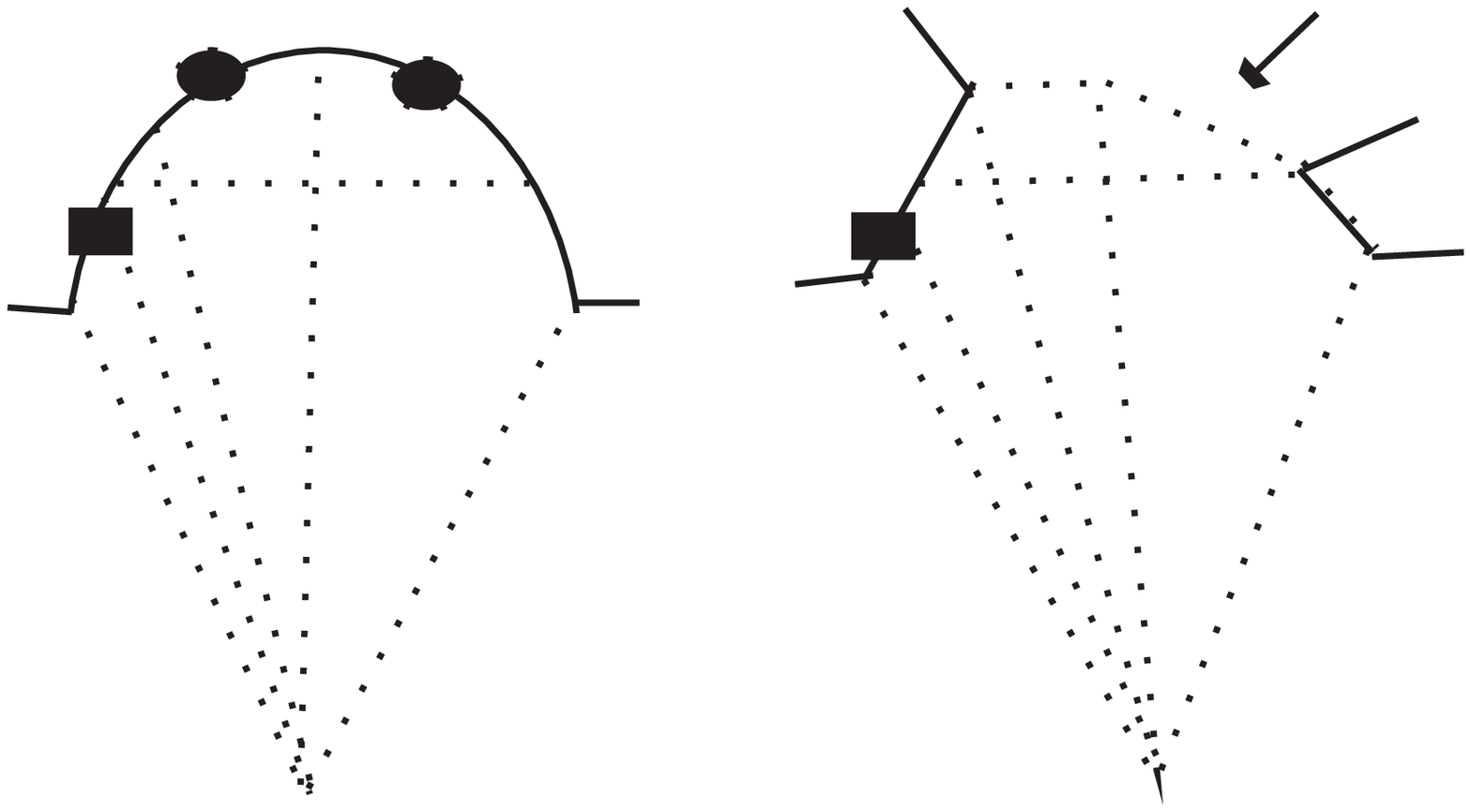}
\caption{\label{figcur}Two Feynman diagrams which result in the same analytic
expression
for nullified external momenta. We expect the one on the lhs
to fulfil the 4TR, while the one on the rhs fail should fail.}
\end{figure}

Consider Fig.(\ref{figcur}).
It shows two Feynman diagrams. The first one contributes
to a correction of a $\bar{\psi}\psi \phi^2$ coupling, say
(it is as well a contribution to $\bar{\psi}\psi \phi^3$,
by making the Yukawa coupling at the top into
a $\bar{\psi}\psi \phi^2$ coupling with a zero-momentum external
boson),
while the other one contributes to $(\bar{\psi}\psi)^2$, for
example. For the graph on the lhs external bosons
couple with zero momentum at the two black blobs, which effectively
bosonizes the two fermion fields as the coupling is of Yukawa type,
with $(1/\kslash){\bf 1}(1/\kslash)=1/k^2$.
Thus, for fermions with vanishing external momentum, both graphs
result in the same analytic expression.\footnote{Once we have
written down a theory involving Yukawa couplings and
non-renormalizable $\phi$-selfcouplings, the other couplings
in both graphs like $\bar{\psi}\psi \phi^2$ and
$(\bar{\psi}\psi)^2$will be
dynamically generated. This is important. When we restrict ourselves
to renormalizable $\phi^4$ selfcouplings, we will not generate any of
the non-renormalizable coupling needed to generate the degeneracy
of Feynman graphs coinciding with the same analytic expression.}

Now we are in a very interesting situation: the analytic expression
we have written down refers to two different Feynman graphs,
and hence could refer to any linear combination of them.
For the graph on the lhs all steps in our proof of the 4TR
go through, while for the graph on the rhs they do not.
If we transport the rectangle
which corresponds to the endpoint $z$ of the propagator
$\Delta_F(x_0-z)$ along the horizontal propagator to the
other three endpoints demanded by the 4TR, it is clear that
the graph on the rhs is in trouble: the horizontal
propagator couples to two internal fermions on the left, while it
couples to one internal fermion and one internal boson on the right.
The four terms for the graph on the rhs thus obey different powercounting
amongst them. Or, from the viewpoint
of proviso $iii)$, we have a situation
$\Delta_{F,1}(s)-\Delta_{F,2}(s)\not=0$,
where we note that the two vertex factors are the same
(they provide pure numbers, coupling constants)
but the propagators for $x_r-x_c$, ($\Delta_{F,1}(s)$), and
$x_d-x_r$, ($\Delta_{F,2}(s)$),  are different.
The arrow in Fig.(\ref{figcur}) denotes the point where
our 4TR will fail. In our notation, it is the graph ${\cal G}_3$.

\begin{figure}
\epsfysize=3cm\epsfbox{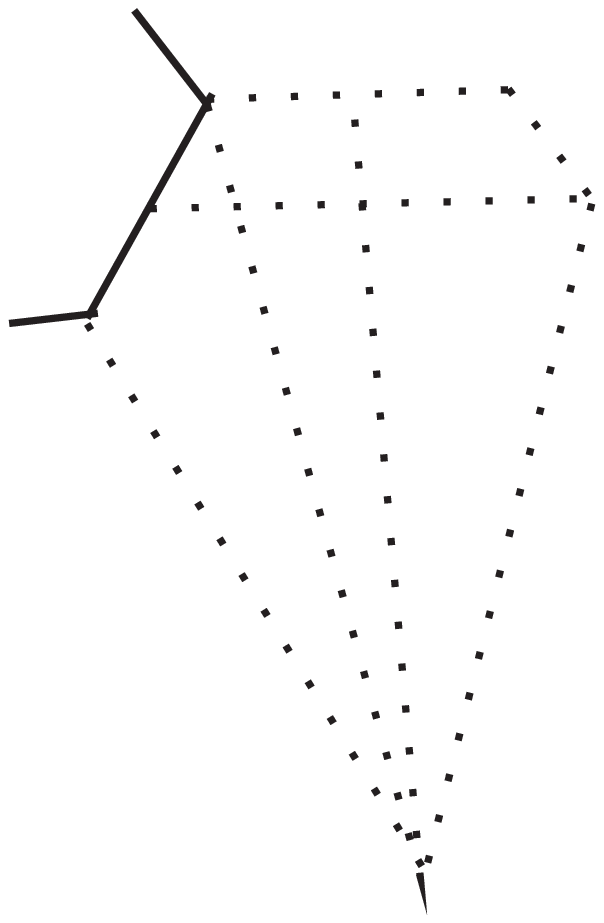}
\caption{\label{fignot}Shrinking the propagator $\Delta_F(x_m-x_l)$
of
Fig.(\ref{figcur}) to a point results
in a five-loop graph with a four-loop subdivergence given
in this figure.}
\end{figure}

Now let us investigate what happens
at the subdivergence, which appears when the propagator for
$x_r-x_m$ shrinks to a point, and whose presence
we believe to be the sole
source of trouble for the graph on the rhs of Fig.(\ref{figcur}).
The corresponding Feynman graph
for this subdivergence is given in Fig.(\ref{fignot}).

We now consider solely the graphs which appear
when the rectangular blob in Fig.(\ref{figcur}) is located on the ``right''
in the graph on the lhs in the figure (in the
figure, we explicitly give it only in the ``left'' ${\cal G}_1$ position).
Then, it sits in the ${\cal G}_3$ or ${\cal G}_4$
position.

According to our previous analysis, we expect
\begin{equation}
0=<I_4+I_5+I_{3,4}+I_{8,4}>.
\end{equation}

Again collecting the remaining divergent contributions,
and using the parts of Props.(1,2,3,4) which are still
valid, we obtain
\begin{equation}
0=<I_{3,4}+I_{5,2}+I_{6,2}+I_{8,4}>.\label{predic}
\end{equation}
Following our reasoning, this should be finite in the
limit $\rho\to 0$ but the presence of the subdivergence.

For $\rho\not=0$, actually only $I_{3,4}$ involves a subdivergence,
due to the presence of the two-point vertex between
$x_r$ and $x_c$ in $I_{8,4}$.
Thus, the divergent behaviour of $
<I_{5,2}+I_{6,2}>$, we claim,  is isolated
in the subdivergence in $I_{3,4}$:
\begin{equation}
<I_{5,2}+I_{6,2}+I_{3,4}>=0.\label{STU}
\end{equation}

So, we only have to consider the limit
\begin{equation}
\lim_{\rho\to 0} I_{3,4}.
\end{equation}
We note that at $u=0$ in $f_4$ the integrand for
$I_{3,4}$ equals a Feynman integral which is identical with
${\cal G}_4$.\footnote{This was first observed by D.J.Broadhurst.}
Integrating the convergent one-loop integral first, and then taking the
limit $\rho\to 0$, one confirms that\linebreak
$<I_{3,4}>=<I_{5,2}>$.
According to Prop.(1) horizontal paths should cancel,
and thus $I_{5,2}=-I_{5,4}\Rightarrow I_{5,4}+I_{3,4}=0$,
in agreement with Prop.(2).

Collecting everything, we end up with the following expectation:
\begin{equation}
<{\cal G}_3-2{\cal G}_4>=0.
\end{equation}
In \cite{BK4} we will report on a calculation confirming these
expectations.

\section{Conclusions}
This finishes our consideration for the moment. We hope that these results
shed some light on the connection between knot theory and UV divergences
in a perturbative quantum field theory.

We also expect that the 4TR
might be of practical value in future calculations, and hope that the
close relation of UV divergences to weight systems, which comes quite
unexpectedly, stimulates field theory, knot theory as well as number theory,
where the question of classification of multiple zeta values (MZVs)
\cite{DZ} remains one of the most fascinating topics.

Also, relation Eq.(\ref{STU}) bears some resemblance with the
STU relation, and deserves further study.

\section*{Acknowledgements}
Ever since I started working on knots and renormalization
I profitted much from David Broadhurst's enthusiasm and support.
I also very much like to thank David Broadhurst for concerned
and detailed reading and corrections
of a preliminary version of the manuscript.

I thank Bob Delbourgo for generous
and encouraging advice and for hospitality
during a stay at the University of Tasmania, where some of these
ideas came to mind.
With equal pleasure I thank Peter Jarvis and Ioannis Tsohantjis
for their stimulating interest in the subject.

\end{document}